\documentclass[iop,numberedappendix]{emulateapj}
\usepackage{natbib}
\usepackage{enumitem}
\newcommand{\msol}{\,\textrm{M}_\sun}                % Solar mass
\newcommand{\ATD}{ATLAS${}^{\rm3D}$}
\shorttitle{ASSESSING CONSISTENCY OF IMF CONSTRAINTS IN THE NEAREST STRONG LENSES}
\shortauthors{ }
\setlist[enumerate]{noitemsep}
\submitted{Accepted to ApJ}
\begin{document}

\title{The Initial Mass Function in the Nearest Strong Lenses from SNELLS:\\
Assessing the Consistency of Lensing, Dynamical, and Spectroscopic Constraints}
\author{Andrew B. Newman$^1$, Russell J. Smith$^2$, Charlie Conroy$^3$, Alexa Villaume$^4$, and Pieter van Dokkum$^5$}
\affil{$^1$ The Observatories of the Carnegie Institution for Science, Pasadena, CA, USA; \href{mailto:anewman@obs.carnegiescience.edu}{anewman@obs.carnegiescience.edu}}
\affil{$^2$ Centre for Extragalactic Astronomy, University of Durham, South Road, Durham, United Kingdom}
\affil{$^3$ Department of Astronomy, Harvard University, Cambridge, MA, USA}
\affil{$^4$ Department of Astronomy and Astrophysics, University of California, Santa Cruz, CA, USA}
\affil{$^5$ Department of Astrophysical Sciences, Yale University, New Haven, CT, USA}

\begin{abstract}
We present new observations of the three nearest early-type galaxy (ETG) strong lenses discovered in the SINFONI Nearby Elliptical Lens Locator Survey (SNELLS). Based on their lensing masses, these ETGs were inferred to have a stellar initial mass function (IMF) consistent with that of the Milky Way, not the bottom-heavy IMF that has been reported as typical for high-$\sigma$ ETGs based on lensing, dynamical, and stellar population synthesis techniques. We use these unique systems to test the consistency of IMF estimates derived from different methods. We first estimate the stellar $M_*/L$ using lensing and stellar dynamics. We then fit high-quality optical spectra of the lenses using an updated version of the stellar population synthesis models developed by Conroy \& van Dokkum. When examined individually, we find good agreement among these methods for one galaxy. The other two galaxies show 2-3$\sigma$ tension with lensing estimates, depending on the dark matter contribution, when considering IMFs that extend to $0.08\msol$. Allowing a variable low-mass cutoff or a nonparametric form of the IMF reduces the tension among the IMF estimates to $<2\sigma$. There is moderate evidence for a reduced number of low-mass stars in the SNELLS spectra, but no such evidence in a composite spectrum of matched-$\sigma$ ETGs drawn from the SDSS. Such variation in the form of the IMF at low stellar masses ($m\lesssim0.3\msol$), if present, could reconcile lensing/dynamical and spectroscopic IMF estimates for the SNELLS lenses and account for their lighter $M_*/L$ relative to the mean matched-$\sigma$ ETG. We provide the spectra used in this study to facilitate future comparisons.
\end{abstract}


\section{Introduction}

The stellar initial mass function (IMF) is both fundamental as a basic outcome of the star formation process and a critical ingredient for interpreting numerous extragalactic observations. Although the IMF is consistent with being independent of environment within the Milky Way \citep[e.g.,][]{Bastian10}, there is little theoretical reason for this to hold across all star-forming regions where the physical conditions can vary widely \citep[e.g.,][]{Hennebelle08,Krumholz11,Hopkins12,Hopkins13,Chabrier14}. Since only the brightest stars beyond the local volume can be individually detected, constraints on the IMF in external systems are challenging and must rely on integrated light observations.

Two such approaches have been applied to early-type galaxies (ETGs). The first is to model the total mass distribution using gravitational lensing, stellar kinematics, or both in combination \citep{Treu10,Auger10,Spiniello11,Thomas11,ATLAS_XX,Barnabe13,ConroyDutton13,Newman13b,Newman13a}. Since the radial distributions and shapes of the dark matter and stars are thought to be distinct, it is possible to separate the components with the aid of a few assumptions. The principal assumptions are that the stellar mass density follows the luminosity density and that the dark matter profile follows a parameterized form, such as a power law or an NFW halo \citep[Navarro-Frenk-White,][]{Navarro96}. This leads to an integral constraint on the IMF, i.e., $M_*/L$. A promising alternative route to $M_*/L$ is to measure the ``graininess'' of the gravitational potential via microlensing-induced flux anomalies, which \citet{Schechter14} have demonstrated using ten multiply imaged quasars.

The second method is based on detailed analysis of absorption line spectra. Recent stellar population synthesis (SPS) models have opened up studies of the chemical abundance patterns and the IMF in unresolved old populations at a remarkable level of detail \citep{CvD12a,CvD12b,vDC12,LaBarbera13,Spiniello14}. While varying the age and metallicity imparts large and well-known changes to integrated galaxy spectra, variations in the IMF also affect surface gravity-sensitive absorption lines at levels that are subtle but detectable ($\simeq1\%$) with high-quality data.

When these two methods are applied to ETGs, both are consistent with a trend of ``heavier'' IMFs in higher-$\sigma$ \citep{ATLAS_XX,LaBarbera15}, more metal-rich \citep{MartinNavarro15_califa}, or more $\alpha$-enhanced \citep{CvD12b} galaxies. (Since these quantities are mutually correlated, the primary driver of the correlation is more difficult to establish.) The ``heaviness'' of the IMF can be expressed via the mass factor $\alpha = (M_*/L)/(M_*/L)_{\rm MW}$, where $(M_*/L)_{\rm MW}$ is inferred from SPS models assuming a fiducial IMF and thereby normalizes differences in age or metallicity. Lensing and dynamical methods infer higher $\alpha$ in high-$\sigma$ ETGs, but cannot distinguish between an increased number of low-mass stars versus stellar remnants, while absorption line studies of such galaxies directly point to a larger number of low-mass ($<1 \msol$) stars, i.e., a ``bottom-heavy'' IMF. 

This apparent convergence of entirely independent techniques has lent credence to claims of a non-universal IMF. Nonetheless, although the global IMF trends may appear consistent, \citet{Smith14} pointed out that some puzzles remain: the IMF constraints obtained for individual galaxies using different techniques do not always correlate well, and the principal galaxy property that underlies the IMF trends is not consistently inferred. Such comparisons depend on the data set and methods used;  \citet{Lyubenova16} found a stronger correlation between total dynamical mass and SPS-based mass.

Such discrepancies could indicate fundamental problems in the methods used to infer the IMF. Alternatively, they could arise from difficulties in directly comparing measurements (e.g., mismatched apertures in galaxies that may have radial IMF gradients, or different IMF parameterizations used in SPS modeling), or they could point to the presence of a second parameter driving IMF variations. Distinguishing these possibilities requires detailed comparisons of IMF constraints from complementary methods in a sample of ``benchmark'' galaxies with high-quality data.

Strong-lensing galaxies are particularly useful for this purpose, since with high-resolution imaging the total mass $M_{\rm Ein}$ projected within the Einstein radius can be measured to 1-2\% precision \citep{Treu10ARAA}. Large samples of galaxy-scale lenses at $z \sim 0.2$ \citep{SLACS_IX,SLACS_X,Treu10} have been used to study the IMF, but lower-redshift lenses are especially valuable for two reasons. First, the Einstein radius is typically located at a smaller fraction of the effective radius, where the dark matter fraction is smaller. Second, obtaining spectroscopy with adequate depth and wavelength coverage to constrain the IMF is more practical for nearby lenses. 

\citet{Smith13} investigated ESO325-G004 (also known as SNL-0), then the closest known strong lens, and showed that it does not follow the mean IMF trends described above. Despite being a massive ETG with $\sigma\sim300$~km~s${}^{-1}$ and ${\rm [Mg/Fe]}\sim+0.3$, its IMF was found to be consistent with that of the Milky Way, with a mass factor $\alpha = 1.04 \pm 0.15$ that is a factor of $\simeq 1.5$ smaller than the mean $\sigma-\alpha$ relations found by \citet{Treu10}, \citet{CvD12b}, and \citet{Cappellari13}. Following this initial study, \citet[][hereafter S15]{Smith15} embarked on the SINFONI Nearby Elliptical Lens Locator Survey (SNELLS), which uncovered two new nearby strong lenses at $z=0.03$ and 0.05. Surprisingly, these two lenses were also shown to favor a relatively ``lightweight'' IMF. The mean $\langle \alpha \rangle = 1.10 \pm 0.08 \pm 0.10$ (statistical and systematic uncertainties, respectively) for the SNELLS sample is consistent with a Milky Way-like IMF and strongly inconsistent with the ``heavy'' IMFs favored by other studies for typical ETGs of similar $\sigma$ or [Mg/Fe].

Do the SNELLS lenses reveal a fundamental inconsistency in different methods used to infer the IMF? Are they indicative of a large scatter in the IMF, perhaps driven by yet unknown parameters? Or can the apparent discrepancies be resolved by relaxing some of the typical assumptions, e.g., allowing more flexibility in the shape of the IMF? In this paper we address these questions by comparing the lensing, dynamical, and spectroscopic methods of IMF estimation for the three SNELLS lenses. We first present lensing and stellar dynamical estimates of the total $M/L$ and stellar $M_*/L$. We then analyze new spectroscopic observations with an updated version of the \citet[][hereafter CvD12]{CvD12b} SPS models and compare these to the lensing and dynamical results.

Throughout we adopt the distances, effective radii, and Einstein radii of the SNELLS lenses measured by S15.

\section{Observations}

\subsection{Magellan/IMACS Spectroscopy}
\label{sec:imacs}

The SNELLS lenses were observed using the IMACS spectrograph \citep{Dressler11} at the 6.5m Magellan Baade telescope during 2015 April 9--10 and 2015 September 25. Spectroscopic observations were made using both the 600/8.6${}^{\circ}$ and 600/13.0${}^{\circ}$ gratings with the f/4 camera in order to cover the wavelength range 3565-9415~\AA~continuously with a uniform resolution of 2.8~\AA. The GG495 filter provided order blocking for the red setup. The $1''$-wide long slit was oriented east--west for SNL-0 and SNL-2 and north--south for SNL-1. Total exposure times ranged from 60~minutes to 100~minutes per grating.

Reduction was performed with a combination of IRAF and custom IDL scripts. The two-dimensional wavelength solution was first obtained from arc lamp exposures. This mapping was then used to model the quartz lamp spectrum and slit illumination profile, which were divided from the dome flats to isolate pixel-to-pixel variations. Dome flats at Magellan are obtained using a screen inserted at the pupil. For observations with the red grating, dome flats were interspersed through the observations to mitigate shifts in fringes induced by flexure as the instrument rotates. These proved to be very minimal. The slit illumination was measured using twilight sky exposures. Following flat fielding, cosmic rays were identified and interpolated using {\tt LACOSMIC} \citep{vanDokkum01}.

For exposures obtained with the red grating, consecutive dithered images were subtracted to obtain a first-pass sky subtraction. The spectra were then rectified, making small zero-point corrections to the wavelength solution using night sky lines. Residual sky emission in the red setup, and all sky emission in the blue setup, was removed by subtracting the flux averaged within apertures extending $33''$-$44''$ from the galaxy center.\footnote{At these distances, the galaxy surface brightness is only $\simeq0.2\%$ of its average value with the extraction aperture described below, so contamination is negligible.}

For our dynamical analysis, we then registered and averaged the two-dimensional spectra from each exposure. For our stellar population analysis, we extracted one-dimensional spectra from each exposure and averaged them. This extraction was weighted along the spatial axis such that the resulting spectrum mimics one obtained within a circular aperture with a $2\farcs2$ radius. Specifically, we applied a weight of unity to the central $1''\times1''$ and weighted the outer rows by the ratio of the area of a half-annulus ($\pi R \times dR$) to the part subtended by the slit ($dR \times s$), or $\pi R / s$, where $s=1''$ is the slit width, $dR=0\farcs22$ is the binned pixel scale, and $R$ is the distance to the galaxy center. This $2\farcs2$ aperture is close to $\theta_{\rm Ein}$ and so enables a direct comparison between lensing and spectroscopic IMF estimates. (We quantify the possible effect of small residual aperture mismatch in Section~\ref{sec:specconstraints}.) This is also the maximum radius for which we could extract a near-infrared spectrum with matched aperture, owing to the shorter slit of the FIRE spectrograph (Section~\ref{sec:fire}).

The spectra were then divided by a flux calibration curve obtained using a combination of the quartz lamp spectrum, which captures high-frequency features in the instrumental response, and observations of white dwarf standards from \citet{Moehler14}. Telluric absorption was removed by fitting synthetic transmission spectra generated by the {\tt molecfit} radiative transfer code \citep{Smette15,Kausch15}. The correction proceeded iteratively. A first-pass SPS model of the galaxy spectrum was fit using {\tt ppxf} \citep{Cappellari04} with the regions of strongest telluric absorption masked. Dividing the galaxy spectrum by this model produces an approximate transmission spectrum, which was then fitted using {\tt molecfit}. The original galaxy spectrum was then divided by this fit synthetic transmission spectrum and refit with SPS models (now without additional masking). After three iterations this process accurately removed telluric absorption over the wavelength range used in our analysis ($\lambda_{\rm rest} < 8900$~\AA). Typical rms residuals were only $1.4\times$ the formal noise; these residuals include random noise along with any systematic errors in the telluric absorption correction and the modeling of the galaxy spectrum.

IMACS disperses onto four detectors, and our spectra therefore have several $\sim50$~\AA~gaps. The final processing step was to resample the stacked spectra from the multiple detectors and multiple gratings into a single spectrum with 100 km~s${}^{-1}$ bins. Small velocity corrections were made to ensure that spectra obtained with the blue and red configurations share a common systemic redshift. The median signal-to-noise ratio per 100~km~s${}^{-1}$ bin redward of Mg~b~was 320 for SNL-0 and SNL-1 and 140 for SNL-2 (equivalently, 210 and 90 per \AA, respectively).

\subsection{Magellan/FIRE Spectroscopy}
\label{sec:fire}

All SNELLS lenses were also observed using FIRE, a near-infrared echellete spectrograph at the Magellan Baade telescope \citep{Simcoe13}, during the nights of 2015 April 8, May 3, and September 25. The FIRE spectra cover $0.82-2.51\mu$m, but in this paper we use only the region around the Wing-Ford band of FeH near 9916~\AA~for SNL-0 and SNL-1. Observations of SNL-2 were also made but had insufficient signal-to-noise ratio and so are excluded from our analysis. On-target exposure times for SNL-0 and SNL-1 were 32~minutes and 54~minutes, respectively. We operated FIRE in the up-the-ramp sampling mode. The $1''$ wide slit provided a resolution of $R\simeq4000$.

The short $6''$ FIRE slit does not permit sky subtraction within the science exposures. Instead, we nodded the telescope $2'$ to obtain a blank sky spectrum every 4~minutes. Initial reduction steps, including wavelength calibration and flat fielding, were performed with the FIREHOSE pipeline. Remaining steps were performed with custom IDL routines. We first modeled and subtracted the background measured between orders, in order to remove scattered light and amplifier offsets. After rectifying the images and improving the wavelength calibration, the sky exposure associated with each sky frame was then modestly rescaled and shifted in wavelength to minimize sky line residuals in each order. Spectra were then extracted in each order, using the same aperture and weighting as the IMACS spectra, and flux calibrated using a white dwarf standard. Orders were finally combined, allowing for small residual flux offsets between orders determined using the regions of overlap. Telluric absorption was removed as described in Section~\ref{sec:imacs}.

We found that the continuum shape varied somewhat in observations made on different nights. Before combining the one-dimensional spectra from multiple nights, we normalized the spectra in the rest-frame 9600-10200~\AA~range using a cubic polynomial and resampled to 100 km~s${}^{-1}$ bins. The final signal-to-noise ratio per 100~km~s${}^{-1}$ bin at the Wing-Ford band is 210 for SNL-0 and 320 for SNL-1 (equivalently, 110 and 170 per \AA, respectively).

\subsection{VLT/X-shooter Spectroscopy}

We acquired optical and near-infrared spectra for all the SNELLS lenses with X-shooter \citep{Vernet11} at the 8.2m UT2 of the ESO Very Large Telescope (VLT). The observations were made using the image-slicing integral field unit (IFU), providing a 4.0$\times$1.8\,arcsec$^2$ aperture, oriented close to the parallactic angle. For each galaxy the data comprise three on-source and two off-source sky exposures ($40''$ distant) observed in an ABABA sequence.

Two dichroics in X-shooter divide the light among three arms. The analyses reported in this paper use only the UVB and VIS data, for which the total exposure time was 23~minutes. The data were processed using the standard X-shooter pipeline to generate five separate data cubes, for each echelle order, from the exposures. Subsequently, we combined the object and sky exposures for each order, and used the ESO {\tt skycorr} \citep{Noll14} and {\tt molecfit} tools to account for sky variation between the exposures, and to derive corrections for telluric absorption. These steps were performed separately for each of the three IFU slices, to help account for small shifts in the wavelength solution residuals between the slices. The echelle orders were matched together using low-order polynomial corrections derived through comparison to a nominal SSP model spectrum. Spectra were extracted by summing the flux within the entire IFU field. Outlying flux points were identified at the original high spectral resolution ($R$\,$\approx$\,8000) before resampling the combined spectra onto 100\,km\,s$^{-1}$ velocity bins.

In comparisons of the X-shooter spectra with the IMACS data, SPS models, and Sloan Digital Sky Survey (SDSS) composite spectra, we noticed a significant ``ringing'' near the wavelength of the UVB-VIS dichroic, which could not be removed by normal data reduction steps. The ringing probably arises from the complex instrument response around the dichroic and its temporal variation \citep[e.g.,][]{Schonebeck14}.

Because of this difficulty, we prefer the IMACS spectra for our main SPS analysis in Section~\ref{sec:spec}. We use the X-shooter data for two purposes: first, to measure the aperture velocity dispersion $\sigma_{e/2}$ in Section~\ref{sec:aperturedispersion}, and second, to fill in a small but important gap in the IMACS spectral coverage around $\lambda_{\rm rest}\approx4160-4260$~\AA. This region contains the Ca4227 feature, which is important to constrain [Ca/Fe]. Specifically, we match the line width (including instrumental resolution and galaxy velocity dispersion), redshift, and continuum shape of the X-shooter spectrum to those of the IMACS spectrum in the vicinity of this gap and then insert the matched spectrum. This procedure assumes that the IMACS and X-shooter apertures probe a similar mean radius within the galaxy. The mean light-weighted radius within the IFU field of view is approximately equal to that within a circular aperture having $R=1\farcs6$. Since this is fairly close to our $R=2\farcs2$ IMACS aperture, we consider that the effect of any abundance gradients will be minimal.

\subsection{Imaging}
\label{sec:imaging}

We obtained $r$-band images of SNL-1 and SNL-2 using the LDSS-3 imaging spectrograph at the Magellan 2 telescope in photometric conditions. Photometric calibration was tied to the SDSS DR9 \citep{Ahn12} using observations of stars in Stripe 82. Night-to-night variations indicate a conservative zero-point uncertainty of $3\%$. For SNL-0, we used Hubble Heritage observations taken with the Advanced Camera for Surveys and the F625W filter (Proposal 10710).
When constructing our dynamical model of SNL-2, we also use an $R$-band image obtained in excellent seeing with FORS2 at the VLT. Since these data were not obtained in photometric conditions, we calibrated by matching the luminosity within $\theta_{\rm Ein}$ to the LDSS-3 measurement (see Section~\ref{sec:lensing}).

\begin{deluxetable*}{lccccl}
\tablecolumns{6}
\tablewidth{0pt}
\tablecaption{Measured Quantities for SNELLS Lenses\label{table}}
\tablehead{\colhead{Quantity} & \colhead{Units} & \colhead{SNL-0} & \colhead{SNL-1} & \colhead{SNL-2} & \colhead{Notes}}
\startdata
\cutinhead{Photometric and spectroscopic properties} \\
Redshift & \ldots & 0.034 & 0.031 & 0.052 &  \\
$R_e$ & arcsec & 9.8 & 3.3 & 5.9 & Effective radius in $J$ band. Denoted $R_{\rm Eff}$ by S15 \\
$\theta_{\rm Ein}$ & arcsec & 2.85 & 2.38 & 2.21 & Einstein radius. Denoted $R_{\rm Ein}$ by S15 \\
$L_{r,\rm Ein}$ & $10^{10} L_{\odot,r}$ & $2.84 \pm 0.06$ & $1.62\pm0.05$ & $2.42\pm0.07$ & $M_{r,\odot} = 4.64$~AB; distance error not included \\
$\sigma_{\rm raw}$ & km s${}^{-1}$ & $335\pm16$ & $290\pm14$ & $274\pm13$ & Measured in $4''\times1\farcs8$ X-shooter IFU aperture \\
$\sigma_{e/2}$ & km s${}^{-1}$ & $312\pm15$ & $289\pm14$ & $263\pm13$ & Aperture correction assumes $R_e$ from S15 \\
\cutinhead{Total $M/L$} \\
$(M/L_r)_{\rm L}$ & $(M/L)_{\odot,r}$ & $5.04\pm0.36$ & $5.49\pm0.42$ & $4.75\pm0.53$ & Total $M_{\rm Ein}/L_{r,\rm Ein}$ with $M_{\rm Ein}$ from S15 \\
$(M/L_r)_{\rm D}$ & $(M/L)_{\odot,r}$ & $5.89\pm0.72$ & \ldots & $5.62\pm0.66$ & Total projected $M/L$ within $\theta_{\rm Ein}$ from \\
&&&&& stellar dynamical modeling only \\
\cutinhead{Stellar $M_*/L$ from lensing/dynamics} \\
$(M_*/L_r)_{\rm L+EAGLE}$ & $(M/L)_{\odot,r}$ & $4.23\pm0.34$ & $4.61\pm0.39$ & $3.51\pm0.46$ & Based on dark matter fractions from EAGLE \\
$(M_*/L_r)_{\rm L+D}$ & $(M/L)_{\odot,r}$ & $4.12\pm0.90$ & \ldots & $4.87\pm0.60$ & From joint lensing and dynamics model \\
\cutinhead{Spectroscopic $M_*/L$} \\
$(M_*/L_r)_{\rm MW}$ & $(M/L)_{\odot,r}$ & $4.04\pm0.10$ & $3.93\pm0.20$ & $3.65\pm0.24$ & Kroupa IMF \\
$(M_*/L_r)_{\rm 1PL}$ & $(M/L)_{\odot,r}$ & $7.58^{+0.69}_{-0.64}$ & $7.80^{+1.01}_{-0.96}$ & $4.96^{+1.04}_{-0.96}$ & Single power-law IMF at $m < 1\msol$ \\
$(M_*/L_r)_{\rm 2PL}$ & $(M/L)_{\odot,r}$ & $7.41^{+0.99}_{-0.86}$ & $7.83^{+1.37}_{-1.14}$ & $4.56^{+1.37}_{-1.00}$ & Double power-law IMF at $m < 1\msol$ \\
$(M_*/L_r)_{\rm 2PL+cut}$ & $(M/L)_{\odot,r}$ & $6.36^{+0.80}_{-0.71}$ & $6.51^{+0.99}_{-0.95}$ & $3.78^{+0.95}_{-0.64}$ & Double power-law with low-mass cutoff \\
$(M_*/L_r)_{\rm non-p}$ & $(M/L)_{\odot,r}$ & $6.24^{+1.12}_{-1.00}$ & $5.57^{+0.59}_{-0.53}$ & $3.82^{+0.61}_{-0.59}$ & Nonparametric IMF \\
\cutinhead{IMF constraints: $\alpha = (M_*/L_r)/(M_*/L_r)_{\rm MW}$} \\
$\alpha_{\rm L+EAGLE}$ & $(M/L)_{\odot,r}$ & $1.05\pm0.09$ & $1.18\pm0.12$ & $0.96\pm0.14$ &  \\
$\alpha_{\rm L+D}$ & $(M/L)_{\odot,r}$ & $1.02\pm0.22$ & \ldots & $1.33\pm0.17$ &  \\
$\alpha_{\rm L+no~DM}$ & $(M/L)_{\odot,r}$ & $1.25\pm0.09$ & $1.40\pm0.13$ & $1.30\pm0.17$ & Total lensing mass; assumes no dark matter \\
$\alpha_{\rm D+no~DM}$ & $(M/L)_{\odot,r}$ & $1.46\pm0.18$ & \ldots & $1.54\pm0.21$ & Total dynamical mass; assumes no dark matter \\
$\alpha_{\rm 1PL}$ & $(M/L)_{\odot,r}$ & $1.87\pm0.18$ & $2.00^{+0.32}_{-0.28}$ & $1.33^{+0.33}_{-0.29}$ & Single power-law IMF at $m < 1\msol$ \\
$\alpha_{\rm 2PL}$ & $(M/L)_{\odot,r}$ & $1.84\pm0.23$ & $1.99^{+0.34}_{-0.29}$ & $1.25^{+0.38}_{-0.28}$ & Double power-law IMF at $m < 1\msol$ \\
$\alpha_{\rm 2PL+cut}$ & $(M/L)_{\odot,r}$ & $1.58^{+0.20}_{-0.18}$ & $1.66^{+0.27}_{-0.24}$ & $1.05^{+0.26}_{-0.20}$ & Double power-law with low-mass cutoff \\
$\alpha_{\rm non-p}$ & $(M/L)_{\odot,r}$ & $1.54^{+0.28}_{-0.25}$ & $1.41\pm0.15$ & $1.04\pm0.17$ & Nonparametric IMF
\enddata
%\tablecomments{}
\end{deluxetable*}

\section{Lensing $M/L$ and $M_*/L$}
\label{sec:lensing}

S15 measured the luminosities $L_{\rm Ein}$ of the SNELLS lenses within their Einstein radii $\theta_{\rm Ein}$ in the $J$ band. We have remeasured $L_{\rm Ein}$ in the SDSS $r$ band in order to facilitate comparisons with other studies.
We used the images introduced in Section~\ref{sec:imaging} to measure the flux within a circular aperture with radius $\theta_{\rm Ein}$ for each lens. The raw flux was then corrected for bandpass shifting and filter differences ($0-0.06$~mag), estimated using FSPS \citep{FSPS1,FSPS2,FSPS3}; Galactic extinction ($A_r=0.05-0.13$~mag; \citealt{Schlafly11}); and the point spread function ($0-0.07$~mag) following S15. The derived luminosities are listed in Table~\ref{table}, with 2-3\% uncertainties estimated from photometric zero-point and Galactic extinction errors. Also listed are the lensing-based total $M/L_r$ based on the masses $M_{\rm Ein}$ found by S15. The lensing $M/L$ is proportional to distance, which was derived from the Hubble flow, and we include an uncertainty corresponding to a 500~km~s${}^{-1}$ peculiar velocity (3-4\%).

SNL-2 has a neighboring galaxy separated by $7''$ that could plausibly contribute to $L_{\rm Ein}$. By fitting two-dimensional S\'{e}rsic models to the galaxies with {\tt galfit} \citep{Peng02}, we estimate that this contribution is only $1.5\%$, a minor effect that we neglect.

Proceeding from the total $M/L$ to the stellar $M_*/L$ requires an estimate of the dark matter contribution within $\theta_{\rm Ein}$. This cannot be estimated from the lensing constraints alone. One route is to appeal to simulations of galaxy formation. Here we follow S15, who used the EAGLE simulations \citep{Schaye15,Schaller15} to estimate dark matter contributions that range from 16--25\% of $M_{\rm Ein}$. Table~1 lists the stellar $(M_*/L_r)_{\rm L+EAGLE}$ derived from lensing and the EAGLE simulations. In the following section, we present an alternate method of estimating the dark matter contribution.

\section{Dynamical $M/L$ and $M_*/L$}
\label{sec:dynamics}

In addition to their utility for modeling the stellar populations of the SNELLS lenses, the IMACS spectra provide the opportunity to measure dynamical $M/L$ and to estimate the stellar $M_*/L$ through a joint lensing+dynamics analysis. Here we describe our kinematic measurements and the dynamical modeling.

\subsection{Stellar Kinematics}
\label{sec:kinematics}
Resolved stellar kinematics were measured using IMACS long-slit spectra. We preferred the IMACS data for this purpose since they cover a larger radial extent than the X-shooter IFU. Spectra were extracted from the combined two-dimensional spectrum for each lens, with the bin size adjusted to reach a minimum signal-to-noise ratio. We used our model fit to the integrated spectrum, described in Section~\ref{sec:spec}, as the template. This template is redshifted, broadened by a Gaussian line-of-sight velocity distribution, multiplied by a 5th-degree polynomial, and added to a 14th-degree polynomial to fit each bin using the {\tt ppxf} code. The procedure accounts for the instrumental and template resolutions.

Reasonable variations in the polynomial order affect the derived $\sigma$ by $\simeq 3\%$, which was added in quadrature to the random errors. Additionally, templates obtained using stars from the MILES library \citep{MILES} or the \citet{Vazdekis12} SPS models lead to dispersions that are uniformly larger than our default values by up to 5\%, which we account for below as a correlated systematic uncertainty. This is likely a conservative estimate of the uncertainty, given that our default template produces a significantly lower $\chi^2$ than the others. 

\begin{figure*}
\centering
\includegraphics[width=0.9\linewidth]{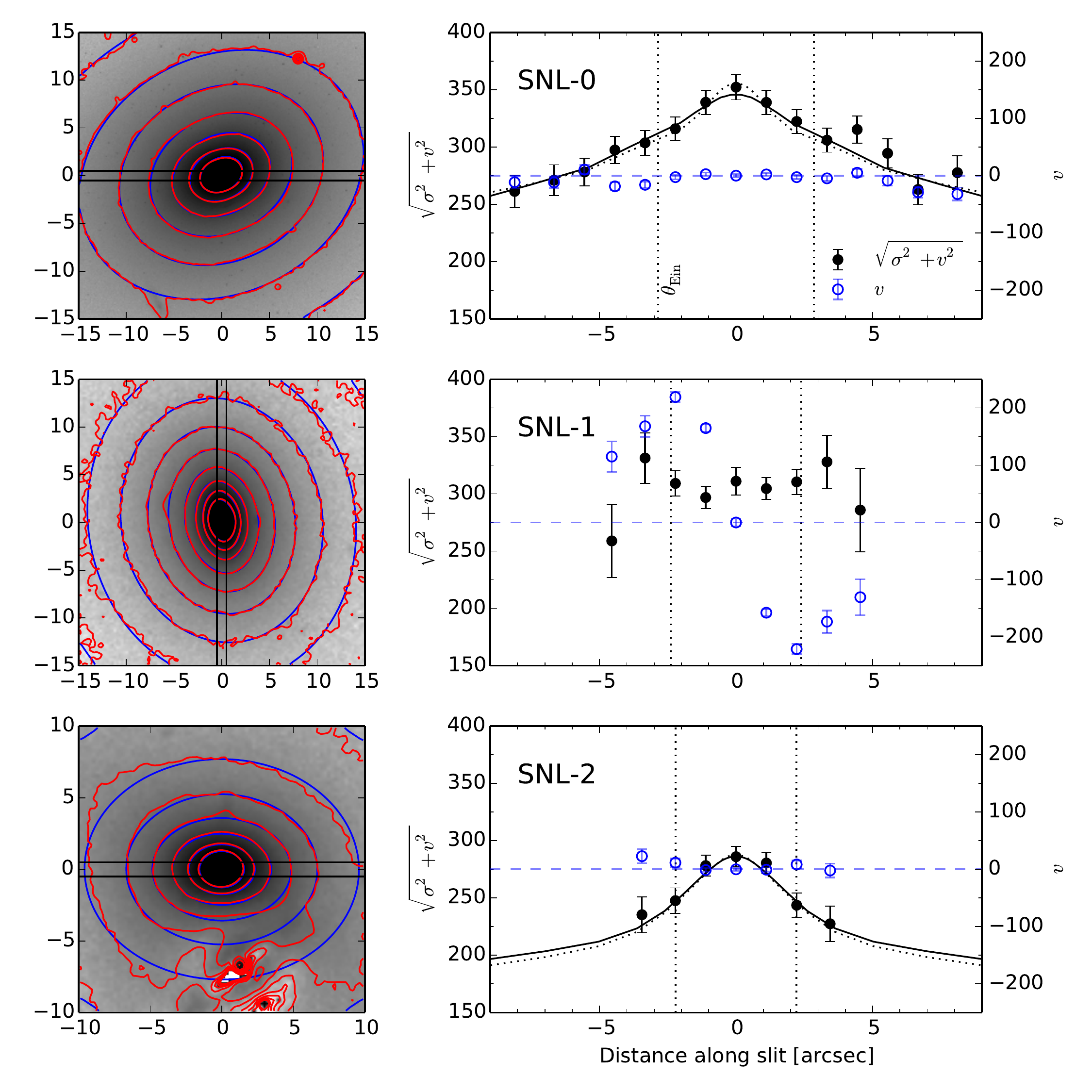}
\caption{\emph{Left:} Isophotal contours (red) are overlaid on the images of each SNELLS lens introduced in Section~\ref{sec:imaging}. Blue contours show the corresponding isophotes of the multi-Gaussian expansion used in our dynamical modeling. Black lines show the position of the IMACS slit. North is up and east is left; axis units are arcsec. \emph{Right:} Observed stellar kinematics along the IMACS slit are shown with the JAM fits overlaid for SNL-0 and SNL-2 (see Section~\ref{sec:dynamicalmodeling}). The solid line is the median posterior model in fits to the dynamics only; the dotted line refers to fits that also include $M_{\rm Ein}$ as a constraint. Note the different axis scales for $\sqrt{\sigma^2+v^2}$ and $v$. Error bars on $\sigma$ do not include an overall 5\% systematic uncertainty. Vertical lines indicate the location of the Einstein radius.\label{fig:kinematics}}
\end{figure*}

The derived kinematic profiles are shown in Figure~\ref{fig:kinematics}. SNL-0 and SNL-2 are entirely pressure supported with no detectable rotation. The situation is different for SNL-1, where rotational support is significant. The IMACS long-slit data do not sample the the velocity field azimuthally, and the X-shooter IFU data are not radially extended enough to constrain the more complicated dynamical structure of SNL-1. For these reasons, we exclude SNL-1 from the dynamical analysis and focus on SNL-0 and SNL-2.

\subsection{$M/L$ from Dynamics Only}
\label{sec:dynamicalmodeling}

Stellar dynamics were calculated using Jeans anisotropic modeling (JAM; \citealt{Cappellari08}). JAM requires that the projected luminosity density be expressed via multi-gaussian expansions, which we show in the left panels of Figure~\ref{fig:kinematics}. SNL-0 and SNL-1 are accurately described by these expansions. Although SNL-2 has more complex isophotes near its neighboring galaxy, which is modeled and subtracted in Figure~\ref{fig:kinematics}, it too is well described by the model within the $R \lesssim 3''$ extent of our kinematic constraints.

Our fiducial mass model consists of three components: (1) the stellar mass traced by the the $r$-band luminosity, (2) a dark matter halo, and (3) a black hole on the \citet{McConnell13} $M_{\rm BH}$-$\sigma$ relation. Given the nearly round ($b/a \simeq 0.8$) isophotes of SNL-0 and SNL-2, by default we use spherical dynamical models. We include a radially invariant velocity anisotropy parameter $\beta_r$ and adopt a prior $\beta_r = 0 \pm 0.3$ that encodes previous findings \citep[e.g.,][] {Cappellari07} that the central stellar kinematics are not very far from isotropic in massive ellipticals. The halo is parameterized with a generalized NFW (i.e., a gNFW) dark matter halo (see Equation 2 of \citealt{Newman13a}). The scale radius is fixed to $r_s = 30$~kpc, following \citet{Treu10} and \citet{Posacki15}. Our constraints are located well within $r_s$, so the density profile is effectively a power-law $\rho_{\rm DM} \propto r^{-\alpha}$. The NFW profile has $\alpha_{\rm DM} = 1$, but it may be modified by the growth of the stellar component. The dark matter density slope in galaxy-scale lenses is not well known, and empirical estimates range from $\alpha_{\rm DM}=1.0-1.7$ (see \citealt{Newman15}, Figure 15). We therefore adopt a uniform prior on $\alpha_{\rm DM}$ over this range. We parameterize the normalization using the fraction $f_{\rm DM}$ of dark matter within a cylinder of radius $\theta_{\rm Ein}$ and take a uniform prior over $[0, 1]$. The free parameters are then $M_*/L$, $\beta_r$, $f_{\rm DM}$ and $\alpha_{\rm DM}$. 

For a given combination of parameters, we generate a map of the projected $\sigma$. This map is blurred by the $1''$ seeing and binned like the data. Since we are using spherical dynamical models, we circularize the radii of the observations.\footnote{The circularized radius is the radius of a circle with equal area to the isophotal ellipse passing through a point.} The parameter space is explored using Markov Chain Monte Carlo (MCMC) sampling. 

To compare to the lensing masses, we computed the total $M/L$ for our model, including all mass components, within a cylinder having a radius $\theta_{\rm Ein}$. We included the distance uncertainty and a 5\% correlated systematic uncertainty in the $\sigma$ measurements. Figure~\ref{fig:kinematics} shows that the JAM models accurately fit the long-slit kinematics. Compared to the total lensing masses within the same aperture, we find that the dynamical estimates of $M/L$ are $\simeq 16\%$ higher for both SNL-0 and SNL-2 at a significance of $\simeq 1\sigma$ (see Table~1).

With the present constraints, consisting only of long-slit kinematics and no higher-order moments of the line-of-sight velocity distribution, there is a degeneracy between $M/L$ and $\beta_r$. When fitting the stellar kinematics alone, we find $\beta_r = 0.17 \pm 0.13$ and $0.13 \pm 0.19$ for SNL-0 and SNL-2, respectively. As we will show below, with higher values of $\beta_r$ it is possible to adequately fit the lensing and dynamics constraints jointly. %It is known that modeling the dynamics of prolate or triaxial galaxies with oblate JAM models can lead to a scatter of 20\% in the inferred $M/L$ \citep{Thomas07,Li16}. Since these geometries are more prevalent at high masses, the assumptions in the dynamical modeling might also be partly responsible for the 16\% differences between the lensing and dynamical $M/L$.

In order to assess systematic uncertainties, we varied several of our model assumptions and inputs to estimate the effect on $M/L$. First, to test for possible gradients in $M/L$ or $M_*/L$ that may be not well described by our mass model, we restricted the velocity dispersion constraints to those within $\theta_{\rm Ein}$. This led to negligible variations of $\simeq5\%$ in $M/L$. Therefore, the dynamical and lensing total $M/L$ refer to apertures that are effectively matched. Second, we considered simpler single-component models in which mass follows light. This led to no change in $M/L$ for SNL-2 and a small increase of 6\% for SNL-0. Third, we varied the mass model components by (1) fixing the dark matter halo to the NFW slope $\alpha_{\rm DM}=1$, or (2) including no black hole or a $10^{10}\msol$ black hole, or (3) considering axisymmetric dynamical models, in which the velocity dispersion ellipsoid is fixed in cylindrical coordinates, rather than the spherical coordinates of our default models. Combinations of these changes produced $M/L_r$ in the range 5.8-6.2 for SNL-0 and 5.1-6.0 for SNL-2, compared to $5.9\pm0.7$ and $5.6\pm0.7$ for the fiducial models, respectively. Thus, the effect of these model variations is comparable to the uncertainty we assigned to the fiducial $M/L$. 

\subsection{$M_*/L$ from Lensing and Dynamics Combined}
\label{sec:LD}

In addition to the total $M/L$, the mass modeling procedure outlined in the previous section allows us to constrain $M_*/L$. This ability to separate the dark and stellar components rests on our chosen prior on the dark matter density profile slope described in Section~\ref{sec:dynamicalmodeling}. Although the separation can, in principle, be done with kinematic constraints alone, the addition of lensing information can help break the degeneracy with velocity anisotropy. Following earlier work on intermediate-redshift lenses \citep[e.g.,][]{Auger10,Spiniello11,Barnabe13,Newman15}, we jointly modeled the stellar kinematics and lensing. We did this using the same mass models described in the previous section, but we now imposed the projected mass $M_{\rm Ein}$ within the Einstein radius (see S15) as an additional constraint. This provides a different route to $M_*/L$ from the lensing and EAGLE simulation-based constraints in Section~\ref{sec:lensing}.

%Considering first the total $M/L$ projected within $\theta_{\rm Ein}$, we find that the lensing+dynamics (L+D) result falls between the dynamics-only and lensing-only values, but with higher inferred $\beta_r$, i.e., more radially biased orbits. Figure~\ref{fig:kinematics} shows that these models are able to fit the kinematics accurately; the $M_{\rm Ein}$ constraint is also satisfied within $1\sigma$. We infer $\beta_r = 0.4\pm0.1$ and $0.2\pm0.2$ for our fiducial models of SNL-0 and SNL-2, respectively. The former value is at the top end of the range reported for massive ellipticals \citep[e.g.,][]{Gerhard01,Cappellari07}. Although this is somewhat concerning, we also find that $\beta_r$ is sensitive to the unknown black hole mass; for example, we infer $\beta_r \approx 0.1$ for SNL-0 if a $10^{10}\msol$ black hole is present.

Figure~\ref{fig:kinematics} shows that the joint models are able to fit the kinematics accurately; the $M_{\rm Ein}$ constraint is also satisfied within $1\sigma$. We infer more radially biased orbits in the joint analysis: $\beta_r = 0.4\pm0.1$ and $0.2\pm0.2$ for SNL-0 and SNL-2, respectively. The former value is at the top end of the range reported for massive ellipticals \citep[e.g.,][]{Gerhard01,Cappellari07}. However, $\beta_r$ is very sensitive to the black hole mass and could be much smaller; for example, we would find $\beta_r \approx 0.1$ for SNL-0 if a $10^{10}\msol$ black hole were present.

Constraints on $(M_*/L_r)_{\rm L+D}$, marginalized over other parameters, are given in Table~1 for the fiducial model: $4.1 \pm 0.9$ and $4.9 \pm 0.6$ for SNL-0 and SNL-2, respectively. These errors include the distance uncertainty and a 5\% correlated uncertainty in the velocity dispersions. We estimated dark matter fractions within the Einstein radius of $f_{\rm DM} = 0.23 \pm 0.10$ for SNL-0 and $f_{\rm DM} < 0.12$ (68\% confidence) for SNL-2. The former is consistent with the EAGLE simulation-based estimate used by S15, while the latter is lower.

% For spherical L+D models: 
%     Same M/Ltot and M/Lstar for SNL-0 regardless of M_BH, but SNL-2 affected by 20% from fiducial to 1e10
% For axisymmetric L+D models:
%     SNL-0: M/Lstar varies 2.8-4.8 from fiducial to 1e10 to fitted 1.9+-0.4 e10
%     SNL-2: M/Lstar varies 3.6-5.4 from fiducial to 1e10 to fitted 1.5+-0.6 e10

To test the sensitivity of $M_*/L$ to the modeling assumptions, we repeated the tests described in Section~\ref{sec:dynamicalmodeling}. For our default spherical dynamical models, the data do not constrain the black hole mass $M_{\rm BH}$. If we assume that a massive $10^{10}\msol$ black hole is present, then $M_*/L$ is reduced from our fiducial models by 3\% and 20\% for SNL-0 and SNL-2, respectively. For the oblate axisymmetric dynamical models, $M_*/L$ is much more sensitive to $M_{\rm BH}$. If we introduce $M_{\rm BH}$ as a free parameter, axisymmetric models for both SNL-0 and SNL-2 favor very massive black holes ($M_{\rm BH} \simeq 2 \times 10^{10} \msol$) and low $M_*/L$ (30\%-40\% below the fiducial values). Such a situation is possible but unlikely, and we consider that more detailed kinematic data with higher spatial resolution and two-dimensional information are required to derive a meaningful constraint on $M_{\rm BH}$. Nonetheless, this exercise shows that $M_*/L$ is more sensitive than the total $M/L$ to the black hole mass and the parameterization of the velocity dispersion ellipsoid. Importantly, the variations to our fiducial dynamical model described above would mainly \emph{reduce} the inferred $M_*/L$.

\subsection{Aperture Velocity Dispersion $\sigma_{e/2}$}
\label{sec:aperturedispersion}

In order to examine trends among ETGs, it is also useful to measure a velocity dispersion within a standardized aperture. We used a circular aperture with radius $R_e/2$. Velocity dispersions $\sigma_{\rm raw}$ were measured from the X-shooter spectra over $\lambda_{\rm rest} \simeq 3700$-8700~\AA, which we preferred for this purpose since the IFU more completely samples the azimuthal distribution, which is complex for SNL-1, than do the long-slit IMACS data. We assumed that $\sigma_{\rm raw}$ probes an effective circular aperture with area equal to the rectangular IFU field. The measured dispersions were then standardized to the $R_e/2$ aperture using the scaling $\sigma(R<R_0) \propto R_0^{-0.06}$ that we derived from high-$\sigma$ galaxies in the \ATD~survey \citep{ATLAS_XV,ATLAS_XX}.
The aperture-corrected $\sigma_{e/2} = \sigma(R<R_e/2)$ are listed in Table~\ref{table}. Although the formal uncertainties are only 1-2~km~s${}^{-1}$, we conservatively assigned 5\% errors based on the template tests described in Section~\ref{sec:dynamicalmodeling}. 

Compared to the velocity dispersions cataloged by the 6dF survey \citep{Campbell14} and used by S15, the present values are systematically lower, even after accounting for aperture differences. This difference probably arises, at least in part, from the Eddington bias: the SNELLS targets were selected to have $\sigma_{\rm 6dF} > 300$~km~s${}^{-1}$, and owing to the steep velocity dispersion function and noisier 6dF data, random scattering will result in biased velocity dispersions for the highest-$\sigma_{\rm 6dF}$ sources. We show in the Appendix that the highest-$\sigma_{\rm 6dF}$ galaxies indeed have 6dF velocity dispersions that are systematically higher than other independent measurements in the literature. We conclude that the discrepancy between our new velocity dispersions and the 6dF values reflects a bias in the 6dF catalog when selecting galaxies in the extreme high-$\sigma_{\rm 6dF}$ tail of the distribution (see Appendix).

\begin{figure*}
\centering
\includegraphics[width=0.55\linewidth]{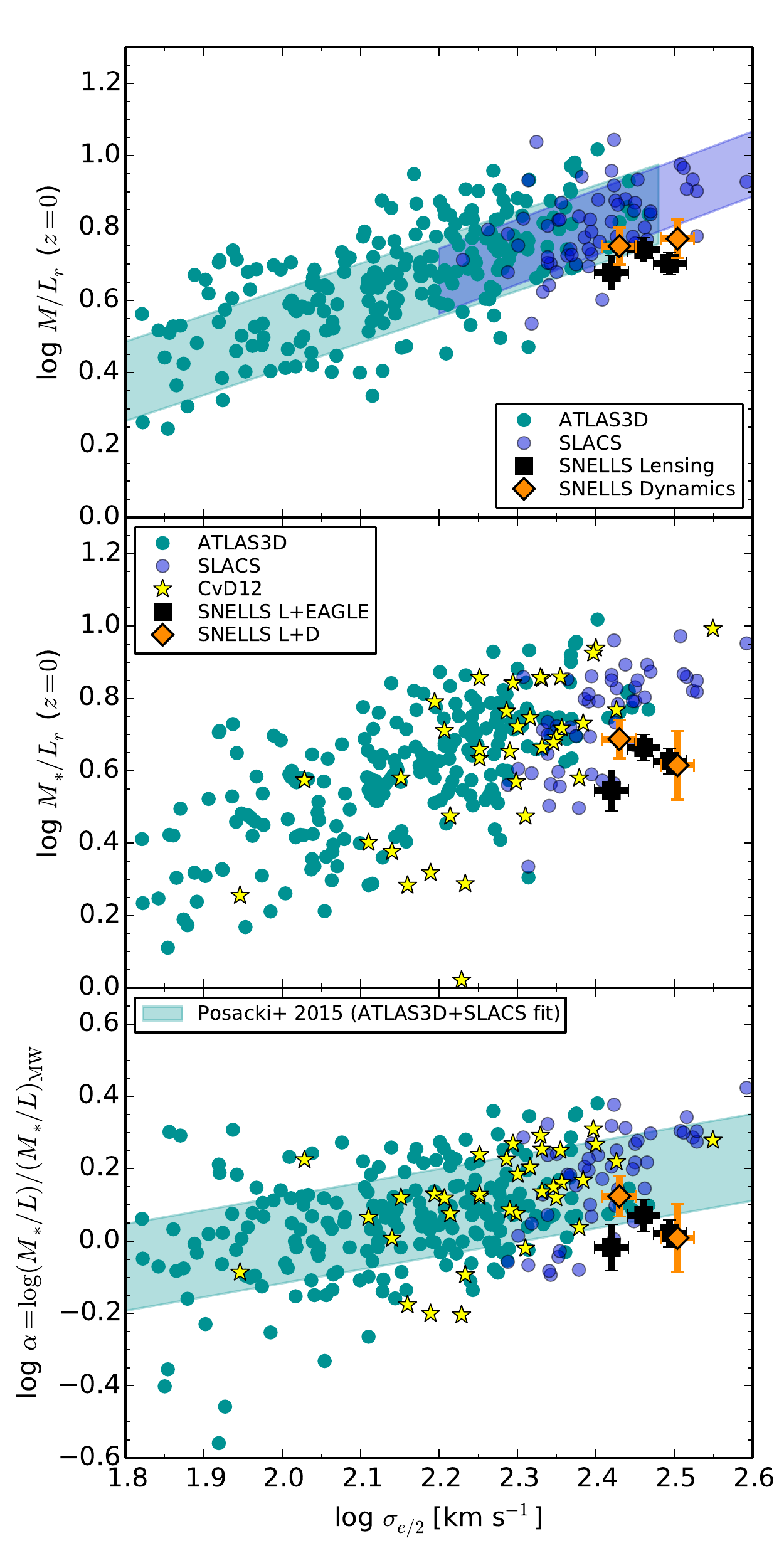}
\caption{\emph{Top panel:} The total $M/L$ of the SNELLS lenses, inferred from lensing or dynamics, are compared to the \ATD~\citep{ATLAS_XV,ATLAS_XX} and SLACS samples \citep{SLACS_IX,SLACS_X} of ETGs. For the latter, we use the $M/L_V$ evolved to $z=0$ by Auger et al.~and convert them into $r$ band assuming a galaxy color $V-r = 0.34$. (This color corresponds to a 10~Gyr old population with $[{\rm Z/H}]=+0.2$ in the FSPS models.) Linear fits from \citet{ATLAS_XV} and \citet{SLACS_X} and their $1\sigma$ scatter are overlaid. \emph{Middle panel:} The stellar $M_*/L$ of the SNELLS lenses, derived from lensing/dynamics, are compared to the same samples and to SPS results from CvD12. The SNELLS results are shown with dark matter contributions from the EAGLE simulations (``L+EAGLE'') or from joint lensing and dynamical modeling (``L+D''). For the SLACS points, we use $M_*/L_r$ from \citet{Posacki15} and approximate the luminosity evolution to $z=0$ using the $V$-band evolution from Auger et al. \emph{Bottom panel}: The IMF mass factor $\alpha$ (Equation~\ref{eqn:alpha}) is compared among the same samples. The SLACS results, and the linear relation from the combined \ATD~and SLACS samples, are from \citet{Posacki15}, with their reported $\alpha$ multiplied by 1.53 to convert from a Salpeter into a Kroupa convention; this factor is derived from the \citet{Vazdekis12} models used by Posacki et al. Note that the SNELLS lenses have low $M/L$, $M_*/L$, and $\alpha$, as estimated from lensing/dynamics, compared to the mean ETG of equal $\sigma$.
\label{fig:compareML}}
\end{figure*}

\section{Results: Lensing and Dynamical Masses and IMF Constraints}
\label{sec:LDresults}

With estimates of $M/L$ and $M_*/L$ derived from our lensing and dynamical analyses, we now consider the SNELLS galaxies in the context of the ETG population and discuss the resulting constraints on their IMF, before comparing these to constraints from SPS models in Section~\ref{sec:spec}.

\subsection{Total $M/L$}
\label{sec:totalML}

As mentioned in Section \ref{sec:dynamicalmodeling}, the total $M/L$ values derived from dynamics alone are 16\% higher than the lensing values for SNL-0 and SNL-2. Although the dynamical values are higher in both cases, which may suggest a systematic difference in the mass estimates, the $\simeq1\sigma$ differences in $M/L$ are also compatible within the uncertainties. Another potential source of error is the luminosity. Since the lensing and dynamical analyses rely on the same images, any error in the photometric calibration would affect both measurements. As we will show below, we obtain lensing-based IMF constraints that are extremely close to those of S15. Since these analyses use the same masses but different images to derive $L_{\rm Ein}$, this rules out significant errors in the luminosities. 

Figure~\ref{fig:compareML} also presents a comparison between the SNELLS sample and the general ETG population. Even before proceeding to the decomposition of dark matter and stars or constraints on the IMF, it is clear that the SNELLS lenses have a uniformly low \emph{total} $M/L$ for their $\sigma$, as compared to relations derived from the \ATD~and SLACS surveys. This is true for both lensing and dynamical estimates, and it suggests the influence of some selection effect in the SNELLS sample. The presence of strong lensing largely depends on $\sigma$ with a cross section $\propto \sigma^4$, so a pure lensing selection should not bias $M/L$ at a given $\sigma$. Studies of the SLACS sample have shown that lenses are not distinguishable from non-lensing ETGs in their photometric, dynamical, or environmental properties, once the samples are matched in $\sigma$ \citep{Bolton08,Treu06,Treu09}. Indeed, lensing (SLACS) and non-lensing (\ATD) samples define very consistent $\sigma-M/L$ relations, and the SNELLS lenses lie below both (Figure~\ref{fig:compareML}, and see \citealt{Posacki15}). The lensing properties of the SNELLS galaxies are also not unusual: $\langle \sigma_{e/2} / \sigma_{\rm SIE} \rangle = 0.97$, consistent with the SLACS lenses. Since this quantity maps closely to the mass density profile slope \citep{Treu09}, the mass structure of the SNELLS lenses is also typical.

Although lenses may not have distinct $M/L$ from the ETG population at fixed $\sigma$, a separate possibility is a selection effect in the SNELLS survey. There are two obvious candidates. First, as discussed by S15, the field of view of SINFONI restricts the range of $\theta_{\rm Ein}$ over which multiple images are detectable. Since $\sigma^2$ is proportional to $\theta_{\rm Ein}$ with only 15\% intrinsic scatter (for given source and lens redshifts), and residuals do not correlate with $M/L$ based on studies of the SLACS sample \citep{Bolton08b}, the field of view may affect the $\sigma$ distribution of the SNELLS lenses but should not introduce any significant bias in $M/L$ at fixed $\sigma$. Second, the surface brightness of the lens could conceivably affect the sensitivity to background sources. However, the SNELLS lenses do not have systematically different surface brightnesses, luminosities, or sizes than the parent 6dF sample (see S15).\footnote{The SNELLS galaxies' $R_e$, $L$, and $I_e = L / (2\pi R_e^2)$ span the range of the parent sample defined by $\sigma > 300$~km~s${}^{-1}$. Turning to three-parameter correlations and examining their positions through the 6dF fundamental plane \citep{Magoulas12}, the SNELLS galaxies are marginally displaced to high $I_e$ at fixed $\sigma$ and $R_e$. Such an offset is expected for galaxies with low $M/L$, since $\Delta I_e \propto \Delta (M/L)^{-1} \propto \Delta (L/\sigma^2 R_e)$ at fixed $\sigma$ and $R_e$.}

Thus, we have not been able to identify a likely selection effect that would favor low-$M/L$ systems in the SNELLS sample. One must bear in mind that in this respect, the SNELLS lenses are not fully representative of high-$\sigma$ ETGs. Nevertheless, they provide an excellent opportunity to test the \emph{consistency} of IMF constraints derived by different methods. 

%\subsection{Stellar $M_*/L$}

%To separate the dark and stellar contributions, we have relied on two methods: estimating the dark matter content from the EAGLE simulations, and a joint lensing+dynamics analysis for SNL-0 and SNL-2. The two methods yield quite consistent results for SNL-0, as demonstrated in Table~1 and in the middle panel of Figure~\ref{fig:compareML}. For SNL-2, $(M_*/L)_{\rm L+D}$ is 40\% higher (1.8$\sigma$) than $(M_*/L)_{\rm L+EAGLE}$. This is due both to the higher total $M/L$ found when including dynamical constraints and to the lower inferred dark matter fraction in this system (see Section~\ref{sec:dynamicalmodeling}) compared to the EAGLE simulations. The error budget in the lensing+dynamics method is naturally larger, given its weaker prior on the dark matter distribution. 

\subsection{Stellar $M_*/L$ and IMF Constraints}

The lensing and dynamical data place integral constraints on the IMF via the mass factor $\alpha$, also referred to as the ``IMF mismatch'' factor:
\begin{equation}
\alpha = \frac{M_*/L_r}{(M_*/L_r)_{\rm MW}},
\label{eqn:alpha}
\end{equation}
where $(M_*/L_r)_{\rm MW}$ is inferred from SPS modeling assuming a fiducial \citet{Kroupa01} IMF. 
S15 assumed $10\pm1$~Gyr ages for the SNELLS lenses in their default results. In a separate model, they instead fit the 6dF spectra to measure $(M_*/L_r)_{\rm MW}$ for each lens, and the resulting $\sim17\%$ uncertainty was larger than that in $M_*/L_r$ and would dominate the uncertainty in $\alpha$.

We fit our higher quality IMACS spectra using the SPS models described in Section~\ref{sec:spec}. The SNELLS lenses span a narrow range of $(M_*/L_r)_{\rm MW}= 3.7 - 4.0$ (see Table~1). Since our spectra now constrain $(M_*/L_r)_{\rm MW}$ to 2-7\%, its uncertainty is not a dominant contributor to the error budget for $\alpha$, so there is no longer any need to rely on the assumption of old ages.

Using the lensing mass and EAGLE dark matter contribution, we find $\alpha_{\rm L+EAGLE} = 1.05 \pm 0.09$, $1.18\pm0.12$, and $0.96 \pm 0.14$ for the three SNELLS lenses, respectively (Table 1). Since these values are mutually consistent, we determine an average by multiplying the probability distributions: $\langle \alpha_{\rm L+EAGLE} \rangle = 1.07 \pm 0.06$, in agreement with the default S15 result, but with the random uncertainties reduced due to the more precise $(M_*/L_r)_{\rm MW}$. Such a ``lightweight'' IMF is well below the trends shown in Figure~2, as demonstrated by S15. 

We have also estimated the stellar $M_*/L$ and dark matter fraction via a joint lensing+dynamics analysis for SNL-0 and SNL-2. The $\alpha$ inferred with this method is quite consistent with the lensing+EAGLE estimate for SNL-0 (Figure~\ref{fig:compareML} and Table~1). However, as discussed in Section~\ref{sec:dynamicalmodeling}, the lensing+dynamics analysis favors a lower dark matter fraction for SNL-2. This increases $\alpha$ by 40\% compared to the lensing+EAGLE estimate and places SNL-2 closer to the mean trend of $\alpha$ with velocity dispersion (Figure~\ref{fig:compareML}, lower panel). Thus, the lensing+dynamics method indicates that the low $M/L$ of SNL-0 arises from a lighter IMF compared to the typical ETG with matched $\sigma$, whereas a lower dark matter fraction may partially contribute in SNL-2.

%Thus, the lensing and dynamical methods indicate that the SNELLS lenses' low total $M/L$ arises from differences in their IMF compared to the typical ETG at matched $\sigma$.

\begin{figure*}
\centering
\includegraphics[width=0.99\linewidth]{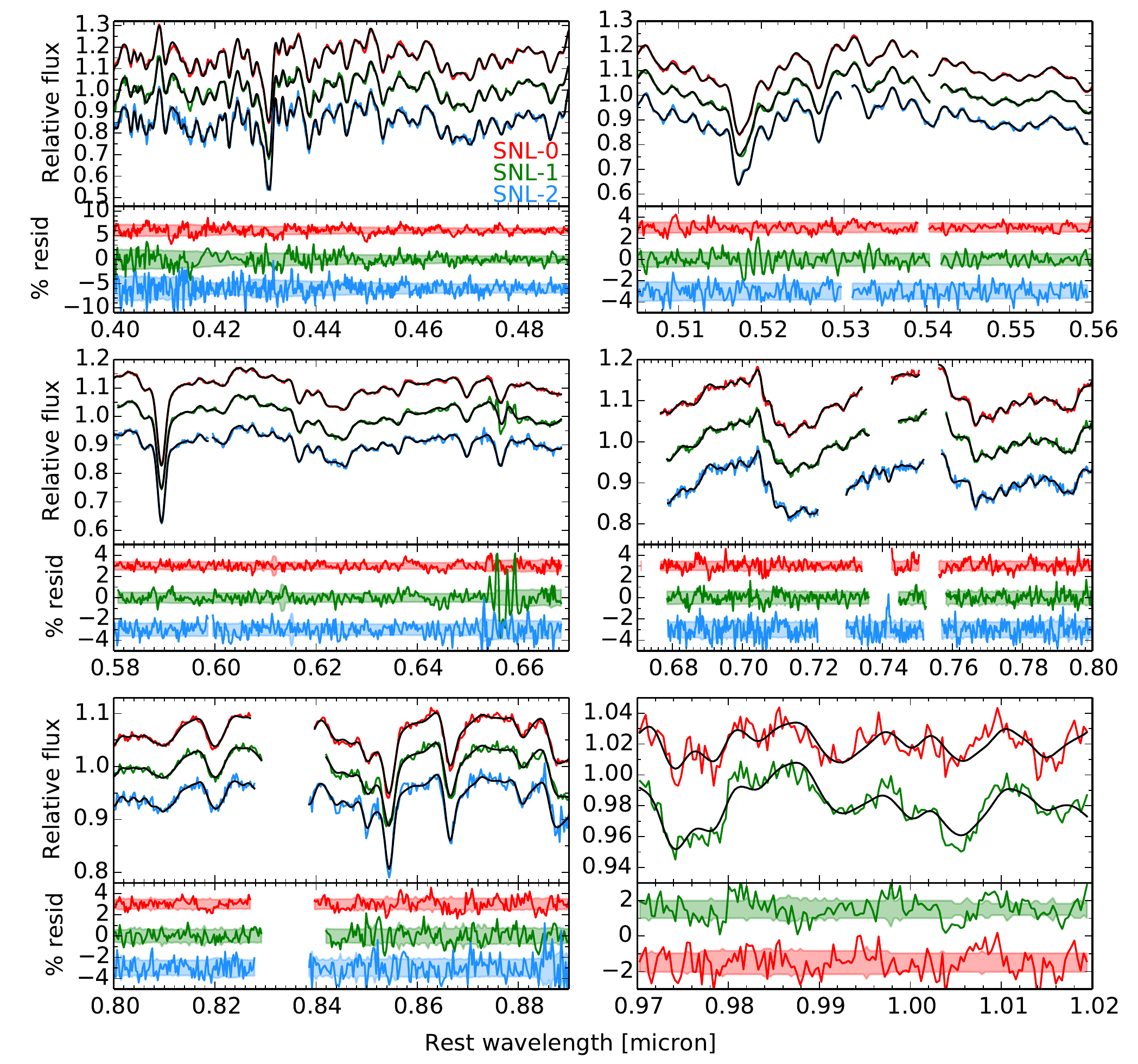}
\caption{Spectra of the SNELLS lenses (colored lines) are compared to the best-fit SPS models (black lines). Residuals are shown in the lower panels, offset for clarity, with the bands encompassing the $\pm1\sigma$ error spectrum. As described in the text, we plot models based on a double power-law IMF, but the fit quality is virtually identical for the other IMF parameterizations.\label{fig:show_bestfits}}
\end{figure*}

\begin{figure*}
\centering
\includegraphics[width=0.8\linewidth]{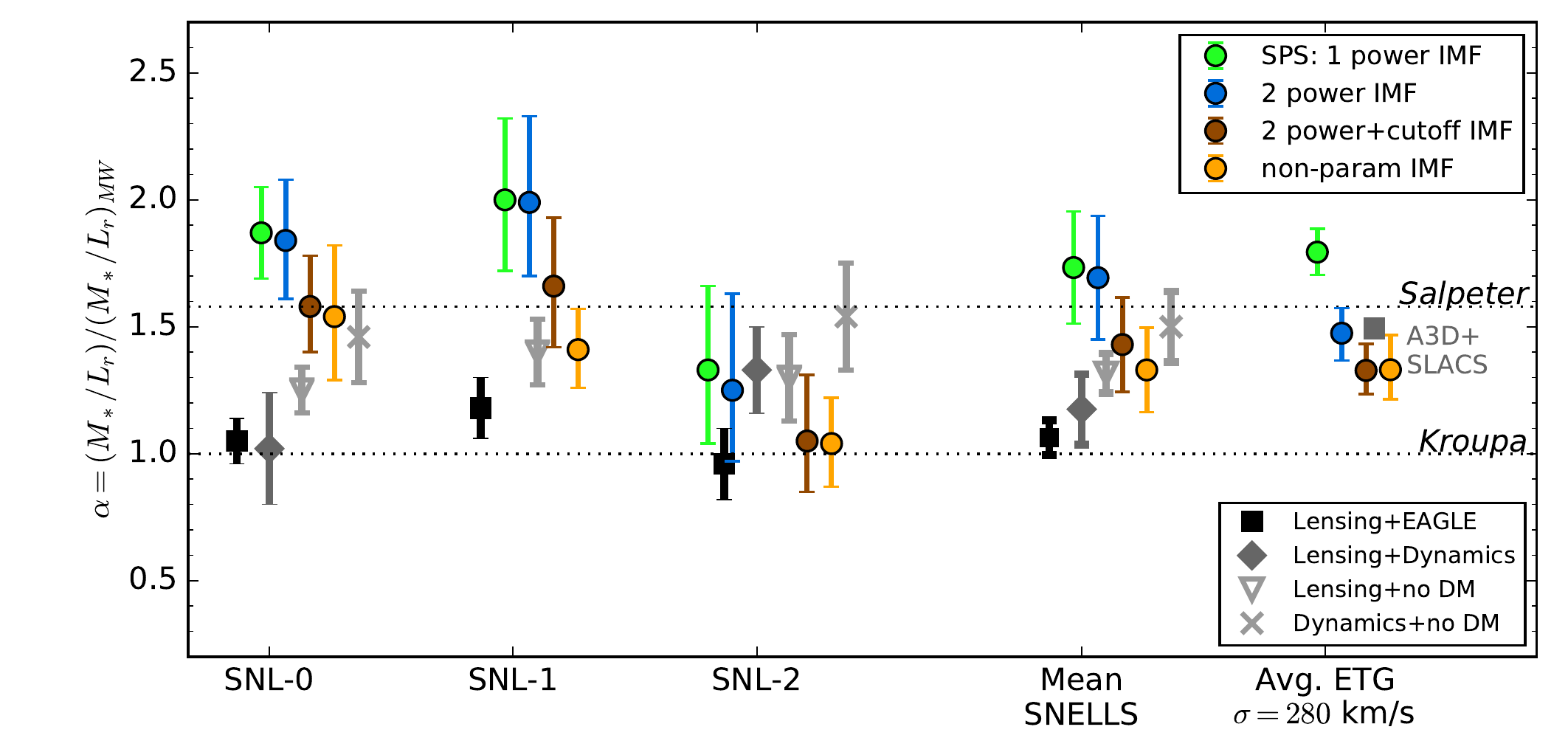}
%\hfill \includegraphics[width=0.375\linewidth]{snells_posteriors_alpha}
%\caption{\emph{Left:} Spectroscopic (colored points) and lensing/dynamical (black/gray points) constraints on the IMF are compared for the SNELLS lenses and an SDSS composite spectrum of ETGs with $\sigma = 280$~km~s${}^{-1}$. The spectroscopic results are shown for the various parameterizations of the IMF indicated in the legend. (Note that the number of power-law segments refers to those below $1\msol$; the slope at $>1\msol$ is fixed as described in the text.) For the SDSS stack, the gray square corresponds to $\alpha$ from the \ATD+SLACS relation constructed by \citet{Posacki15}, evaluated at $\sigma_{e/2} = 280$~km~s${}^{-1}$ and converted from a Salpeter IMF to the reference Kroupa IMF used in this paper. \emph{Right:} Marginalized posterior probability distributions for all $\alpha$ constraints for the SNELLS lenses. Colors have the same meaning as in the left panel.
\caption{Spectroscopic (colored points) and lensing/dynamical (black/gray points) constraints on the IMF are compared for the SNELLS lenses and an SDSS composite spectrum of ETGs with $\sigma = 280$~km~s${}^{-1}$. The spectroscopic results are shown for the various parameterizations of the IMF indicated in the legend. (Note that the number of power-law segments refers to those below $1\msol$; the slope at $>1\msol$ is fixed as described in the text.) For the SDSS stack, the gray square corresponds to $\alpha$ from the \ATD+SLACS relation constructed by \citet{Posacki15}, evaluated at $\sigma_{e/2} = 280$~km~s${}^{-1}$ and converted from a Salpeter IMF to the reference Kroupa IMF used in this paper.
\label{fig:snells_sps}}
\end{figure*}

\section{Results: Spectroscopic IMF Constraints}
\label{sec:spec}

With constraints on the IMF mass factor $\alpha$ now in hand from lensing and stellar dynamics, we consider the results obtained from SPS modeling. We use models that are descended from those presented by CvD12, but which have undergone major changes over the intervening period. These include the use of new isochrones from the MIST project \citep{Choi16}, a new spectral library based on observations from IRTF \citep{Villaume17}, and new age- and metallicity-dependent response functions. The new models span a significantly wider range in age (1-13.5~Gyr) and metallicity ($-1.5 < {\rm [Z/H]} < +0.25$). %The models and fitting techniques, and their application to the general ETG population will be described in forthcoming papers by Conroy et al.~(in prep.).

For the SNELLS analysis, we fit single-age, mono-abundance models with parameters describing (1) the age, redshift, velocity dispersion, and IMF, (2) the abundances of C, N, O(=Ne, S), Na, Mg, Si, K, Ca, Ti, V, Cr, Mn, Fe, Co, Ni, Cu, Sr, Ba, and Eu, (3) nebular emission lines from the Balmer series, [\ion{O}{3}], [\ion{N}{2}], [\ion{S}{2}], and [\ion{N}{1}] with a common redshift and velocity dispersion independent of the stars, (4) contributions from additional hot star light and a ``frosting'' of young stars (0.5-3 Gyr), and (5) residual sky emission and absorption, and a constant rescaling of the error spectrum. We do not include nuisance parameters representing additional M giant light or isochrone temperature shifts, unlike CvD12. Further details are provided by \citet{Conroy17}.

The assumed shape of the IMF can have a significant effect on the derived $M_*/L$. To explore this dependence, we have constructed and fit models with four parameterizations. In all cases we assumed a slope $\xi \propto m^{-2.35}$ for masses $>1\msol$. In order of generality, the IMF parameterizations at lower masses are as follows.
\begin{enumerate}
\item A single power-law over $0.08\msol$-$1\msol$.
\item A double power-law over $0.08\msol$-$1\msol$ with a break at 0.5$\msol$.
\item A double power-law as in (2), but with a truncation mass allowed to vary from $m_{\rm cut} = 0.08$-$0.4\msol$.
\item A nonparametric form in which the IMF weights are interpolated among four bins distributed between $0.08\msol$ and $1\msol$, as described by \citet{Conroy17}.
\end{enumerate}

The parameter space is explored using MCMC techniques. The models are constrained by the spectra over six wavelength ranges: 4000-4900~\AA, 5050-5600~\AA, 5800-6700~\AA, 6700-8000~\AA, 8000-8900~\AA, and 9700-10200~\AA. (Small gaps between the first three regions match gaps in the IMACS spectra.) The latter range includes the Wing-Ford FeH band and is not available for SNL-2 or our analysis of SDSS composite spectra.
%, but we will show that its inclusion does not appreciably affect the inferred $\alpha$.
 Within each wavelength range, a polynomial of order $\Delta \lambda / (100~\rm\AA)$ is allowed to modulate the continuum shape of the models to best match the data.

\subsection{Spectroscopic Constraints for SNELLS Lenses}
\label{sec:specconstraints}

The resulting fits are illustrated in Figure~\ref{fig:show_bestfits} for the case of the double power-law IMF. %The data do not discriminate strongly among the IMF parameterizations. Therefore, although we plot the model with a two-part power-law IMF, its fit quality is virtually indistinguishable from the others.
The quality of fit is very high: rms residuals are 0.6\%, 0.8\%, and 1.1\% for the three lenses, respectively, compared to the rms formal errors of 0.4\%, 0.5\%, and 1.0\% within the fitted regions (per 100~km~s${}^{-1}$ bin).

The spectroscopic constraints on $\alpha$ are summarized in Table~1. In Figure~\ref{fig:snells_sps}, we compare these constraints among the IMF parameterizations and to the lensing and dynamical estimates derived in Section~\ref{sec:LDresults}. The spectroscopic estimates for the single and double power-law models are nearly indistinguishable for the SNELLS lenses. When a cutoff $m_{\rm cut}$ is included in the IMF parameterization or when a more general nonparametric form is used, $M_*/L$ declines by $\simeq 20\%$, as we discuss further in Section~\ref{sec:priors}. We note that \citet{LaBarbera13} and \citet{Lyubenova16} found even larger dependences on the IMF form, but those studies considered a single slope extending from $0.6 \msol$ to beyond $1\msol$, which leads to variations in $M_*/L$ from both remnants and dwarfs.

Given the additional complexities in the reduction of the FIRE spectra and our lack of a near-infrared spectrum of SNL-2, we have tested the sensitivity of our results to the inclusion of the spectral region around the Wing-Ford band. Excluding this region has a negligible effect on the derived $M_*/L$ and $\alpha$ for SNL-0 and SNL-1. Furthermore, motivated by concerns about the accuracy of SPS models in matching the strengths of optical and near-infrared \ion{Na}{1} features \citep[e.g.,][]{Smith15b}, we have also performed fits excluding the line at 8190~\AA. As we show in Appendix~B, this lowers the inferred $\alpha$ by $\simeq 30\%$, a significant effect that we return to when we discuss our interpretation in Section~\ref{sec:disc}. The sensitivity of $\alpha$ to the \ion{Na}{1} 8190~\AA~feature, and its relative insensitivity to the inclusion of the Wing-Ford band, are both consistent with the findings of CvD12 (see their Figure~12).

In general, a major hinderance in comparing IMF estimates is the matching of spatial apertures, which is very important given that the IMF may vary radially \citep{MartinNavarro15_radvar,MartinNavarro15_relic,LaBarbera16,McConnell16,Zieleniewski17,vanDokkum17,Davis17}. A key strength of our SNELLS comparison is that aperture differences are very minimal, since the spectroscopic aperture is close to the Einstein radius.\footnote{In the case of multiple stellar populations with distinct $M_i$ and $L_i$, which can also represent the case of a radial $M/L$ gradient, the spectroscopic method will approximately measure a luminosity-weighted $M/L$, i.e., $\Sigma [(M/L)_i L_i] / \Sigma L_i$. Lensing measures $\Sigma M_i / \Sigma L_i$, which is precisely equivalent.} For SNL-1 and SNL-2 these radii are matched within $\leq 8\%$. For the worst case of SNL-0, where $\theta_{\rm Ein} = 2\farcs85$, we can roughly estimate the effect of an IMF gradient by assuming that it has a similar slope to that inferred by \citet{LaBarbera16} for a single massive lens galaxy. For their measured gradient, our spectroscopic $M_*/L$ should be reduced by 8\% to match the lensing aperture, which is not enough to substantially affect our comparisons. Therefore, the lensing, dynamical, and spectroscopic IMF estimates refer as closely as possible to the same aperture.

%With all lensing, dynamical, and SPS estimates of $M_*/L$ for each galaxy, we can assess the mutual consistency of these techniques. Agreement between the spectroscopic and lensing/dynamical $M_*/L$ is mixed among the SNELLS sample. Taking $(M_*/L)_{\rm L+EAGLE}$ and the two-part power-law IMF as specific points of comparison, the spectroscopic and lensing/dynamical measures differ by $3.4\sigma$, $2.7\sigma$, and $1.0\sigma$ for SNL-0, SNL-1, and SNL-2, respectively. Therefore, SNL-2 is consistent, but there is tension for the other two galaxies, with the spectroscopic $M_*/L$ being significantly higher than the lensing/dynamical measure. The tension is reduced somewhat if alternative dark matter contributions or IMF parameterizations are considered; we develop these comparisons further in Section~\ref{sec:disc}.

With all lensing, dynamical, and SPS estimates of $M_*/L$ for each galaxy, we can start to assess the mutual consistency of these techniques. For SNL-2, Figure~\ref{fig:snells_sps} shows that all methods yield consistent results regardless of the modeling assumptions. For SNL-0 and SNL-1, the results of the consistency test depend on the assumed parameterization of the IMF and the dark matter contribution. We will develop these comparisons in detail in Section~\ref{sec:disc}. First, we develop a comparison sample that will prove useful for interpreting our SNELLS measurements.

\begin{figure*}
\centering
\includegraphics[width=\linewidth]{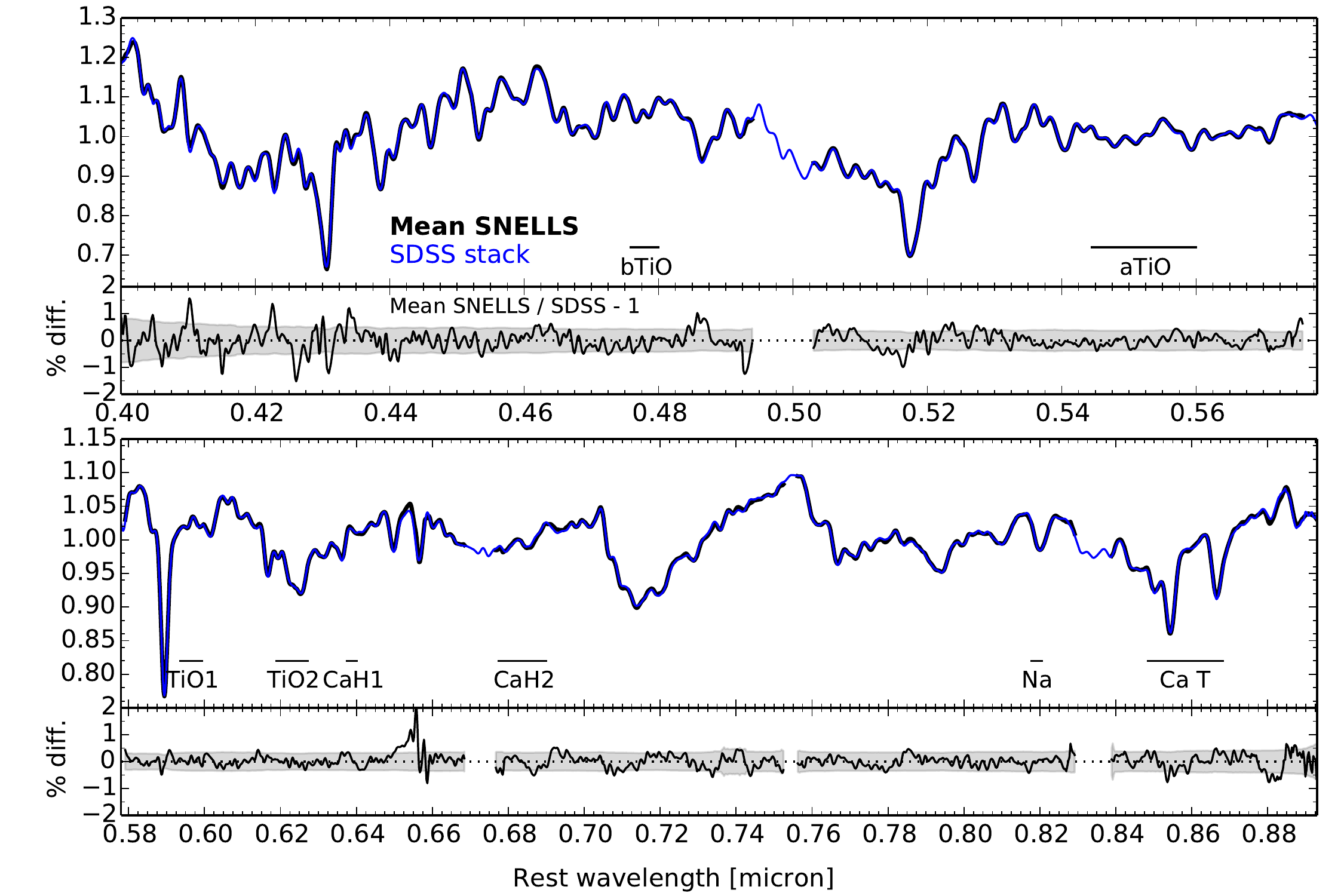}
\caption{The mean spectrum of the SNELLS lenses is compared to our SDSS composite spectrum of ETGs with $\sigma = 280$~km~s${}^{-1}$. The spectra have been matched in line width and continuum shape as described in the text. Residuals are shown in the lower panels with the gray band indicating the $\pm1\sigma$ formal uncertainty. The IMF-sensitive features identified by \citet{Spiniello14} are labeled. Overall, the spectra are remarkably similar; the rms differences are 0.32\% per 100~km~s${}^{-1}$ bin.
\label{fig:comparetosdss}}
\end{figure*}

\begin{figure}
\centering
\includegraphics[width=\linewidth]{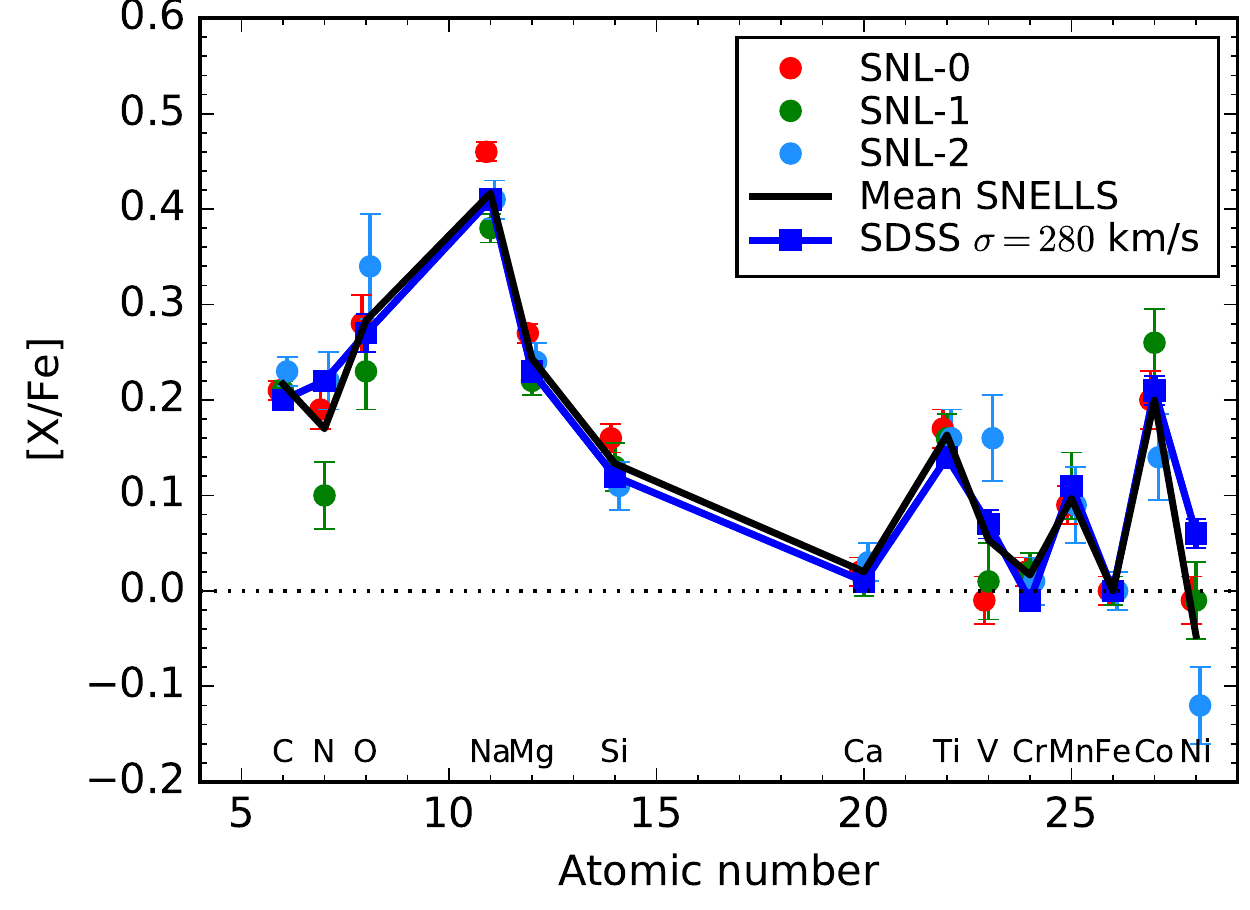}
\caption{Abundance ratios of the SNELLS sample are compared to the SDSS composite spectrum, demonstrating the typical abundance pattern of the SNELLS galaxies. Only elements with well-constrained abundances are plotted.\label{fig:abundances}}
\end{figure}

\subsection{SNELLS Lenses Compared to a Composite ETG\label{sec:sdss_stack}}

All lensing estimates, regardless of the dark matter contribution, indicate that the SNELLS lenses have systematically low $\alpha$ for their velocity dispersions. It is therefore interesting to ask how the spectra and spectroscopically inferred $M_*/L$ of the SNELLS lenses compare with those of an average ETG with similar velocity dispersion.
%: the average offset from the \citet{Posacki15} relation is $\langle \Delta \log \alpha \rangle = -0.19$, with a scatter of only 0.03 based on the $\alpha_{\rm L+EAGLE}$ estimates (see Figure~\ref{fig:compareML}).

For this purpose, we use a composite spectrum constructed from a carefully selected sample of ETGs drawn from the SDSS. ETGs are selected with cuts based on the stellar mass--star formation rate relation and the equivalent widths of H$\alpha$ and [\ion{O}{2}]. The set of stacked spectra will be described and analyzed in a forthcoming paper (C.~Conroy et al. 2017, in preparation). The stacks are created in bins of velocity dispersion that are 0.1~dex wide. For our purposes, we use a spectrum centered on $\sigma=280$~km~s${}^{-1}$, which is very close to the mean of the SNELLS sample, $\langle \sigma_{e/2} \rangle = 288$~km~s${}^{-1}$.

Figure~\ref{fig:comparetosdss} presents a direct comparison between the mean spectrum of the SNELLS lenses and the SDSS stack. We have carefully interpolated the SNELLS spectra to the redshift and sampling of the SDSS stack, convolved them to match the SDSS line width (accounting for both velocity dispersion and instrumental resolution), and warped the SNELLS spectra to match the continuum shape of the SDSS stack before averaging them. Continuum shapes were matched by fitting the ratio spectrum using a cubic spline with knots spaced by 200~\AA. The median signal-to-noise ratio of the SDSS spectrum redward of Mg~b is 370 per 100~km~s${}^{-1}$ (equivalently, 250 per \AA). 

Overall, the mean SNELLS and SDSS spectra are extremely similar and show an rms difference of only 0.32\% (in 100~km~s${}^{-1}$ bins), comparable to the measurement errors. We note that differences around H$\alpha$ and H$\beta$ arise from nebular emission in SNL-1, which is likely driven by the presence of warm gas rather than intrinsic differences in its stellar population. No visually significant differences in the IMF-sensitive regions marked in Figure~\ref{fig:comparetosdss} are evident at this signal-to-noise ratio. Correspondingly, when we fit the SDSS stack using the same models that were applied to the SNELLS lenses, we find that the inferred $\alpha$ is close to the mean of the SNELLS sample for each IMF parameterization (Figure~\ref{fig:snells_sps}, left panel).

As expected from the similarity of the spectra, the abundance patterns of the SNELLS lenses are typical. Figure~\ref{fig:abundances} shows that in the mean, the SNELLS abundance ratios match those of the SDSS stack well. In particular, {[}Na/Fe{]}, {[}Ca/Fe{]}, and {[}Ti/Fe{]} agree very closely, which is relevant since most of the IMF-sensitive features involve these elements. Likewise, approximating the total metallicity as ${\rm [Z/H]} = {\rm [Fe/H]} + 0.9 {\rm [Mg/Fe]}$, we find that the SNELLS lenses have a mean $\langle {\rm [Z/H]} \rangle = 0.26$ that is quite close to the value ${\rm [Z/H]} = 0.24$ measured in the SDSS stack.

\begin{figure*}
\centering
\includegraphics[width=\linewidth]{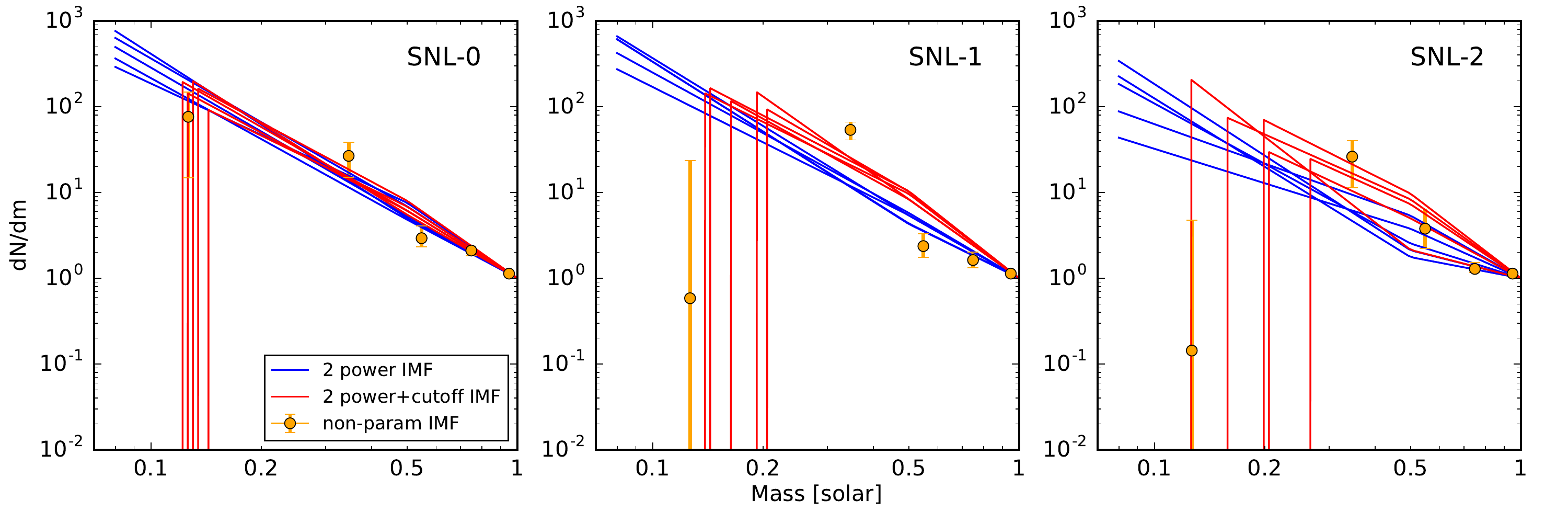}
\caption{Inferred IMFs for the SNELLS lenses. Results are shown for three of the four IMF parameterizations, with the single power-law omitted for clarity. For the double power-law IMFs with (red lines) and without (blue lines) a variable low-mass cutoff, 5 posterior samples are shown to illustrate the uncertainties. For the nonparametric IMF (orange points), the IMF weight and its uncertainty is plotted in each of the mass bins described by \citet{Conroy17}.\label{fig:imf_posteriors}}
\end{figure*}

\begin{figure}
\includegraphics[width=\linewidth]{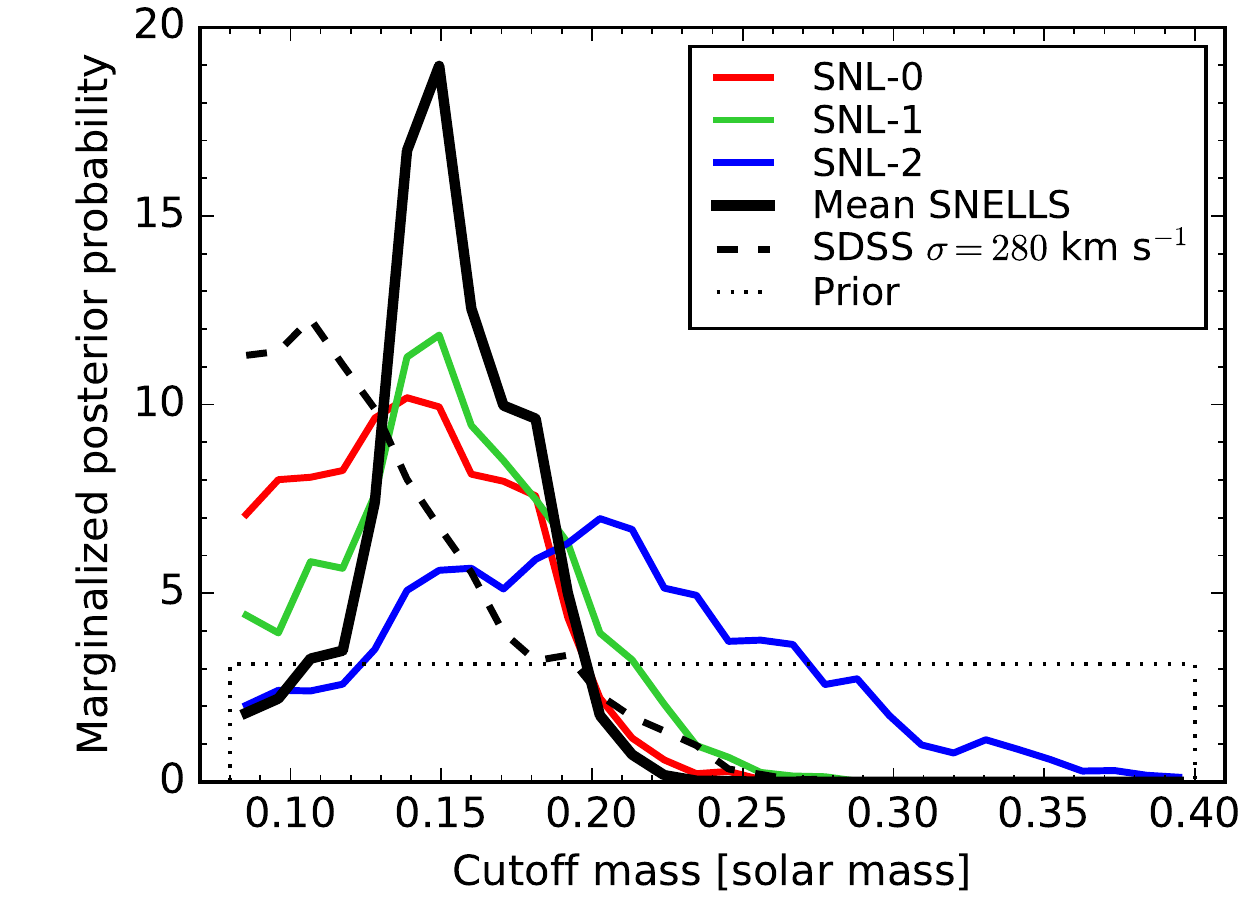}
\caption{Marginalized posterior probability densities for $m_{\rm cut}$, as inferred for the SNELLS lenses and the SDSS stack using the double power-law+cutoff IMF parameterization. The thick line is the probability distribution for the mean $m_{\rm cut}$ of the SNELLS ensemble, assuming that the lenses share a common value, and is obtained as the product of the posteriors of the individual lenses.\label{fig:mcut_posterior}}
\end{figure}

\subsection{The Influence of Priors and IMF Parameterizations on the Inference of $\alpha$\label{sec:priors}}

Our comparison of the masses and the stellar population properties of the SNELLS lenses and the SDSS stack has revealed an apparent contradiction. Lensing indicates a lighter IMF in the SNELLS lenses compared to the typical ETG of matched $\sigma$, as determined by the \ATD~and SLACS surveys. On the other hand, when the spectra are analyzed using the \emph{same} IMF parameterization and priors, the inferred $\alpha$ is always consistent between the SNELLS sample (in the mean) and a composite SDSS spectrum of matched-$\sigma$ ETGs. Furthermore, independent of the stellar population modeling, the spectra themselves are very similar. Here we investigate a possible way to reconcile these observations.% by considering the effects of the priors on the inferred $\alpha$ for our more general IMF parameterizations.

%Figure~\ref{fig:snells_sps} shows a general trend for $\alpha$ to decline as the IMF parameterization becomes more complex. In Figure~ we investigate the origin of this trend by comparing the IMFs inferred for the SNELLS sample under 3 assumed parameterizations. This comparison shows that introducing a variable low-mass cutoff to the double power-law model moves the cutoff to higher masses. Likewise the nonparametric models tend to have less weight in the lowest-mass bin compared to the double power-law model. Both of these lead to lower $M_*/L$ and $\alpha$.

Figure~\ref{fig:imf_posteriors} shows the spectroscopically inferred IMFs for the SNELLS sample. As the complexity of the IMF parameterization is increased, either through the introduction of a low-mass cutoff or a fully nonparametric form, the models favor a smaller contribution of low-mass stars relative to the double power-law model. An important question is the extent to which this smaller contribution is driven by the priors versus the likelihoods (i.e., the data). For example, the prior on $m_{\rm cut}$ in the double power-law+cutoff parameterization is uniform between 0.08$\msol$ and 0.4$\msol$. Thus, $m_{\rm cut}$ can only be higher than the canonical hydrogen-burning limit, and $\alpha$ can only decrease compared to the simpler model with a fixed low-mass cutoff. The decline in $\alpha$ could arise either because the data prefer a cutoff above the hydrogen-burning limit or because the data do not strongly constrain $m_{\rm cut}$. In the latter case, the lower inferred $\alpha$ could be due entirely to marginalizing over $0.08\msol<m_{\rm cut}<0.4\msol$. Likewise, the priors on the IMF weights for the nonparametric model (see \citealt{Conroy17}) are centered on sub-Salpeter slopes, which could drive the inferred $\alpha$ downward (toward the prior) if the data do not strongly constrain the number of low-mass stars.

\begin{figure*}
\centering
\includegraphics[width=0.9\linewidth]{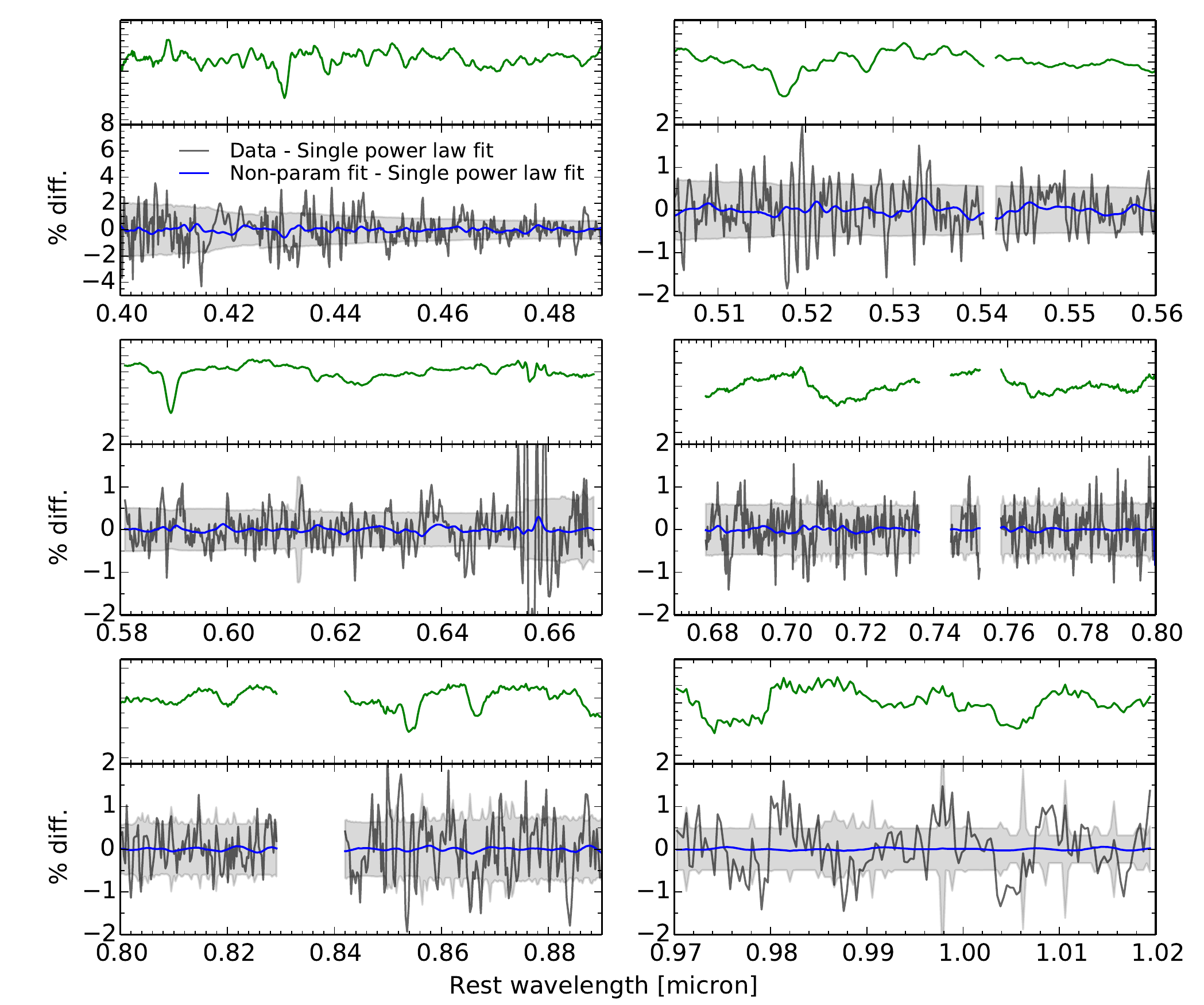}
\caption{Differences between the single power-law IMF and nonparametric IMF model fits to SNL-1 (blue line) are compared to the residuals from the single power-law model (gray) in the lower panels. The shaded region denotes the $\pm1\sigma$ measurement uncertainties. The upper panel in each figure repeats the SNL-1 spectrum from Figure~\protect\ref{fig:show_bestfits}. The difference between the best-fit models are very small---much less than the residuals---despite a difference in the IMF mass factor of $\Delta\alpha=0.6$ between them.\label{fig:compare_modelfits}}
\end{figure*}

%Determining the complexity in the IMF form that the data require is a question of model selection. One method is to compare the evidence, or Bayes factor. However, our MCMC sampling methods---already computationally intensive given the high dimensionality of the parameter space---are optimized for generating posterior samples and not for the even more intensive task of computing the evidence.

A robust model comparison can be made for one particular case of interest, the double power-law IMF parameterizations with and without a variable low-mass cutoff. This comparison is possible because the models are ``nested'': the latter is simply a special case of the former with $m_{\rm cut} = 0.08\msol$. In Figure~\ref{fig:mcut_posterior} we show the marginalized posterior probability distributions for $m_{\rm cut}$ for the SNELLS lenses and the SDSS stack.

The posterior for the SDSS stack peaks near the hydrogen-burning limit; therefore, the decline in $\alpha$ when a variable low-mass cutoff is introduced (compare blue and red points in Figure~\ref{fig:snells_sps}) is not because the data demand $m_{\rm cut} > 0.08\msol$, but instead arises from marginalizing over elevated $m_{\rm cut}$ values that are disfavored but not ruled out. The situation is different for the SNELLS lenses. For each of the lenses, the posterior peaks above 0.08$\msol$, indicating a preference for an elevated cutoff mass, although this preference is not significant for individual systems. The thick line in Figure~\ref{fig:mcut_posterior} shows the product of the posterior probability densities, which is a constraint on the mean $m_{\rm cut}$ for the SNELLS ensemble. The posterior probability density at $m_{\rm cut} = 0.15\msol$ is a factor of 10 higher than that at 0.08$\msol$. This is moderate evidence, but not decisive, in favor of an elevated cutoff mass from the spectra alone.\footnote{The spectra cover the Wing-Ford band for SNL-0 and SNL-1, but not the SDSS stack. We verified that this difference is not the cause of the different posteriors in Figure~\ref{fig:mcut_posterior} by repeating the analysis with the Wing-Ford band masked.}

Is such a difference in the low-mass IMF consistent with the similarity of the mean SNELLS and SDSS spectra demonstrated in Section~\ref{sec:sdss_stack}? In Figure~\ref{fig:compare_modelfits} we focus on SNL-1 and compare the residuals from the best-fitting single power-law model (gray lines) to the difference between that model and the best-fitting nonparametric model (blue lines). These models were chosen because their corresponding IMF mass factors differ by $\Delta\alpha=0.6$, which is the largest dependence on parameterization that we find (see Figure~\ref{fig:snells_sps}). At $\lambda > 5800$~\AA, where most of the IMF constraining power lies, the two models never differ by more than 0.15\% (excluding the H$\alpha$+[\ion{N}{2}] region). This demonstrates that although the mean SNELLS and SDSS spectra differ very little, with an rms scatter of 0.32\%, there are models based on different IMF parameterizations that differ even less (0.15\%) and still have significantly different $\alpha$.

Taken at face value, the small differences between these models can still be detected when aggregated over the full spectrum of a galaxy: comparing these two models, denoted $m_1$ and $m_2$, we find $\Delta \chi^2 = \Sigma_i (m_{1,i} - m_{2,i})^2/\sigma_i^2 = 8$, where the sum is over spectral pixels. However, such small differences can only be detected statistically and are not easily ascribed to individual spectral features (Figure~\ref{fig:compare_modelfits}). Moreover, there are likely systematic errors in the models at the $\sim 0.5\%$ level (e.g., \citealt{vanDokkum17}, Figure~8), which makes the interpretation of such small spectral differences uncertain. This makes it difficult to establish a definitive difference in the low-mass IMF of the SNELLS lenses and the SDSS stack from the spectra alone.
 
In summary, there is moderate evidence for a difference in the low-mass ($m \lesssim 0.3\msol$) IMFs of the SNELLS lenses compared to the SDSS stack. This is not inconsistent with the similarity of their spectra because the model differences can be even smaller. Although the spectra alone cannot decisively establish such a difference, the truncated IMF they suggest in the SNELLS lenses could explain their lower $M_*/L$ (relative to the mean matched-$\sigma$ ETG) inferred using lensing.

\section{Discussion}
\label{sec:disc}
\subsection{Summary and Comparison of IMF Constraints}
We have presented new observations of the three low-redshift ($z=0.03$-0.05) ETG strong lenses located in the SNELLS survey (S15). The presence of strong lensing and the feasibility of very high-quality spectroscopy make them excellent systems for testing the consistency of IMF estimates derived from lensing, stellar dynamics, and stellar population synthesis.

The most fundamental and robust test is the no dark matter limit: does the SPS-based $M_*/L$ exceed the total $M/L$? SNL-2 clearly satisfies this test. For SNL-0 and SNL-1, the results of the test depend on the IMF parameterization. When using a single or double power-law model that extends down to $0.08\msol$,\footnote{We emphasize that contrary to some other usage in the literature, ``single'' and ``double'' refer to the number of power-law segments below $1\msol$; at $>1\msol$ we always use the Salpeter slope.} the spectroscopic $M_*/L$ exceeds the \emph{total} lensing mass. The significance of this difference is 2.5$\sigma$ and $1.9\sigma$ for SNL-0 and SNL-1, respectively, for the double power-law model. Models with more flexibility at the low-mass end of the IMF largely eliminate this tension. For the double power-law model with a variable low-mass cutoff, the differences with respect to the the total lensing masses are reduced to 1.7$\sigma$ and 1.0$\sigma$ for SNL-0 and SNL-1, respectively, while for our nonparametric model these differences are only 1.1$\sigma$ and 0.1$\sigma$. The lensing masses are more precise and less sensitive to the modeling assumptions (e.g., velocity anisotropy) than the dynamical masses. Nonetheless, if we consider the total dynamical $M/L$ for SNL-0, we find that it is also more consistent with the SPS estimates based on the more flexible IMF parameterizations, although the tension with the single and double power-law models is lessened. Altogether, the SPS-based $M_*/L$ does not exceed the \emph{total} lensing or dynamical masses for the SNELLS lenses, if they have a deficit of low-mass stars relative to the double power-law model. This scenario is moderately favored by the spectra (Figure~\ref{fig:mcut_posterior}).

Turning to comparisons of the stellar $M_*/L$, the consistency between lensing/dynamical and SPS estimates depends on the dark matter fraction, which we have evaluated in two ways. First, considering the lensing masses after subtracting a dark matter contribution estimated using the EAGLE simulations, we find that $(M_*/L)_{\rm L+EAGLE}$ and SPS estimates differ by $4.6\sigma$ and $3.1\sigma$ for SNL-0 and SNL-1, respectively, assuming a single power-law IMF; $3.4\sigma$ and $2.7\sigma$ for a double power-law IMF; $2.7\sigma$ and $1.9\sigma$ for a double power-law IMF with a low-mass cut-off; or $1.9\sigma$ and $1.5\sigma$ for the nonparametric IMF model. For SNL-2, all methods of IMF estimation are consistent regardless of the dark matter contribution or IMF parameterization. Notably, all discrepancies are reduced to $<2\sigma$ when considering models based on the nonparametric IMF.

Second, we have independently estimated the dark matter fraction via a joint lensing+dynamical analysis for SNL-0 and SNL-2. The resulting $\alpha$ closely matches the lensing+EAGLE-based estimate for SNL-0, and correspondingly, the dark matter fraction $f_{\rm DM}$ is consistent with the EAGLE estimate and inconsistent with a scenario with no dark matter at the 2.3$\sigma$ level. The main systematic uncertainty arises from the black hole mass and orbital structure, as discussed in Section~\ref{sec:LD}, but variations to our default model mainly shift $M_*/L$ downward and thus would increase tension with the SPS estimates. For SNL-2, although the $\alpha$ inferred from lensing+dynamics is higher than the EAGLE-based estimate, both are consistent with the SPS estimates for all IMF parameterizations.

\subsection{Variations at the Low-mass End of the IMF?}

A key result of this paper is that consistency between lensing, dynamical, and spectroscopic $M_*/L$ estimates requires that in two of the SNELLS lenses, the IMF below 0.5$\msol$ is not a power-law extending down to the canonical hydrogen-burning limit of 0.08$\msol$. Additional flexibility at the low-mass end ($m \lesssim 0.3\msol$) is needed, at least in the context of the CvD models and full-spectrum fitting techniques used here. When this flexibility is provided in the form of our nonparametric IMF, the spectroscopic $M_*/L$ is within $2\sigma$ of $(M_*/L)_{\rm L+EAGLE}$. This mild discrepancy may be reduced if the SNELLS lenses have a smaller dark matter contribution than suggested by simulations, as the lensing+dynamics analysis suggests for SNL-2. Furthermore, since we know that the SNELLS galaxies have an atypical total $M/L$ for their $\sigma$, it is reasonable to suppose that their formation histories and dark matter fractions could also be atypical. 

In addition to better agreement with lensing and dynamical measures, analyses of the spectra alone favor a reduced contribution of low-mass stars, either via an elevated low-mass cutoff $m_{\rm cut} = 0.15 \pm 0.03 \msol$ in models parameterized as such, or via a turn-over in the IMF at $m \lesssim 0.3\msol$ in our nonparametric IMF models. The evidence from the spectra alone is only of moderate significance and not definitive, but it becomes more interesting when coupled with consistent evidence for a lightweight IMF provided by lensing/dynamics. If we accept these results at face value and assume that systematic errors in the methods are not dominant, then this physical scenario appears to be the most plausible one that can reconcile the IMF inferences for all methods.

%In contrast, we find that the typical ETG with matched $\sigma$ does not favor a low-mass cutoff above the hydrogen-burning limit. An analysis of our SDSS composite spectrum produces a posterior distribution of $m_{\rm cut}$ peaking at $0.08\msol$. Further, the spectroscopic $\alpha$ agrees with lensing/dynamical estimates from the \ATD~and SLACS surveys when assuming a double power-law IMF with no elevated cutoff (Figure~\ref{fig:snells_sps}). Imposing a low-mass cutoff above the hydrogen-burning limit is not necessary and slightly harms this agreement. This comparison assumes that \ATD/SLACS surveys and our SDSS stack comprise representative samples of ETGs at a given $\sigma$, which is likely the case, and that they sample comparable apertures; although the latter is not exactly true, the differences are not very large.\footnote{This is particularly true when comparing SLACS, which samples $R_{\rm Ein} \simeq R_e/2$, to our SDSS stack, which samples $\simeq R_e/3$ on average.}

If the SNELLS lenses have a reduced contribution of low-mass stars, this may point to diversity in the low-mass IMF rather than a general property of high-$\sigma$ ETGs (see also \citealt{Leier16}). 
All of our lensing-based estimates, regardless of dark matter fraction, imply that all of the SNELLS lenses have a lighter IMF than the mean matched-$\sigma$ galaxy in the \ATD and SLACS surveys (Figure~\ref{fig:snells_sps}). Only if we adopt the total dynamical mass \emph{and} assume there is no dark matter would the SNELLS lenses' IMF be consistent with the IMF of the mean matched-$\sigma$ galaxy. Considering the limitations of our stellar kinematic constraints and analysis, we judge that the lensing constraints are more robust. They are certainly more precise, and the dynamical measures are consistent with them at the $1\sigma$ level. Improved stellar kinematic data (such as wide-field integral field spectroscopy) and a more sophisticated dynamical analysis (such as Schwarzschild modeling) could provide a yet more stringent test of the consistency of the masses.

While lensing/dynamics measures imply that the SNELLS lenses have atypically light IMFs and therefore that there is scatter in the IMF of high-$\sigma$ ETGs, these techniques cannot elucidate the nature of the such variations. Spectroscopic modeling suggests that variations at low stellar masses ($m \lesssim 0.3\msol$) may be the origin. Our SPS modeling moderately favors a truncated low-mass IMF in the SNELLS lenses, but we find no such evidence when analyzing a stacked spectrum of matched-$\sigma$ galaxies (Figure~\ref{fig:mcut_posterior}). A reduced contribution of low-mass stars appears necessary to reconcile the spectroscopic and lensing/dynamical $\alpha$ estimates for the SNELLS lenses, but it is not necessary to reconcile the spectroscopic $\alpha$ inferred from our SDSS stack with the lensing/dynamical results from the \ATD/SLACS surveys.\footnote{This comparison assumes that \ATD/SLACS surveys and our SDSS stack comprise representative samples of ETGs at a given $\sigma$, which is likely the case, and that they sample comparable apertures; although the latter is not exactly true, the differences are not very large. This is particularly true when comparing SLACS, which samples $R_{\rm Ein} \simeq R_e/2$, to our SDSS stack, which samples $\simeq R_e/3$ on average.} If the SNELLS lenses have an IMF that is deficient at low stellar masses compared to the typical matched-$\sigma$ ETG, this could explain the samples' different lensing/dynamical $\alpha$ while remaining consistent with the SPS constraints.

Diversity at the high-mass end of the IMF affecting the number of remnants could also produce variations in $M_*/L$ that have no direct effect on the spectra. Available constraints on the low-mass X-ray binary (LMXB) populations of ETGs do not favor such diversity \citep{Peacock14,Coulter17}, although the statistics are limited. Varying the number of high-mass stars significantly would also be expected to leave a signature in the abundance pattern, which we do not see in the SNELLS sample (Fig.~6).

\citet{Barnabe13} and \citet{Spiniello15} investigated the low-mass end of the IMF in strong-lensing ETGs at intermediate redshifts. Rather than constraining a low-mass slope and cutoff mass $m_{\rm cut}$ from the spectra and testing consistency with a lensing/dynamical $M_*/L$, as done in this paper, they instead inferred the low-mass slope from the spectra and varied $m_{\rm cut}$ to match the $M_*/L$ from lensing/dynamics. \citet{Barnabe13} inferred $m_{\rm cut} = 0.13\pm0.03$ while \citet{Spiniello15} found $m_{\rm cut} = 0.131^{+0.023}_{-0.026}$. These values lie between the canonical hydrogen-burning limit and our inference for the SNELLS ensemble and are consistent with both. Currently, the uncertainties are too large to discern scatter in the low-mass IMF at high significance.

In general, the stellar population properties of ETGs depend systematically on at least two parameters \citep[e.g.,][]{Graves09}, and the IMF might as well. The puzzle is that the SNELLS lenses do not appear to be systematically different from matched-$\sigma$ ETGs in any parameter except $M/L$ (see S15 and discussion on abundances in Section~\ref{sec:sdss_stack}). \citet{Spiniello15} suggested that the low-mass slope of the IMF might vary with $\sigma^2/R_e^2$ and pointed out that the SNELLS lenses have atypically high values of this parameter, which is proportional to the dynamical mass density, when compared to their XLENS sample. With the revised velocity dispersions presented in this paper, coupled with a correction for the typical wavelength dependence of the effective radius (e.g., \citealt{Vulcani14}, Fig.~16; note S15 measured $R_e$ in the $J$ band), we find that this systematic offset disappears. Thus, if the SNELLS lenses are revealing scatter at the low-mass end of the IMF among high-$\sigma$ galaxies, this does not appear to correlate tightly with galaxy size, luminosity, density, metallicity, or abundance ratios.

%This implies that if scatter in the low-mass IMF is present in high-$\sigma$ ETGs, it will be difficult to firmly establish.
As Figure~\ref{fig:compare_modelfits} shows, SPS models based on different IMF parameterizations with different mass factors $\alpha$ can produce very similar red spectra. If it is possible to measure variations in the low-mass IMF, it will require larger samples of high-$\sigma$ galaxies with both very high-quality spectra covering all of the major IMF-sensitive features and precise masses measured in the same aperture. Surveys such as CALIFA and MaNGA are pushing in this direction.

\begin{deluxetable*}{lccccc}
\tablecolumns{6}
\tablewidth{0pt}
\tablecaption{Spectra of SNELLS Lenses Used in SPS Analysis\label{tab:data}}
\tablehead{\colhead{Galaxy Name} & \colhead{Wavelength (\AA)} & \colhead{$F_{\lambda}$ (arb. units)} & \colhead{$\sigma_{F_{\lambda}}$} & \colhead{Flag} & \colhead{Resolution (km~s${}^{-1}$)}}
\startdata
SNL-0 & 3928.923 & 57.450 & 1.324 & 1 & 89.1\\
SNL-0 & 3930.233 & 56.515 & 1.296 & 1 & 89.1\\
SNL-0 & 3931.545 & 60.762 & 1.302 & 1 & 89.0\\
SNL-0 & 3932.856 & 59.469 & 1.283 & 1 & 89.0\\
SNL-0 & 3934.168 & 57.700 & 1.262 & 1 & 89.0\\
\ldots & \ldots & \ldots & \ldots & \ldots & \ldots
\enddata
\tablecomments{Table~2 is published in its entirety in the machine-readable format. A portion is shown here for guidance regarding its form and content. The spectra were obtained with IMACS and FIRE and extracted to mimic a circular aperture with $R=2\farcs2$. Wavelengths are in vacuum in the observed frame. Pixels excluded in our fits have a flag of 0. The instrumental resolution is provided in terms of the Gaussian $\sigma$.}
\end{deluxetable*}

\subsection{Other Possibilities}

We have argued that scatter in the form the IMF at $m \lesssim 0.3\msol$ is a plausible way to reconcile all IMF estimates for the SNELLS lenses and to account for their lighter IMFs compared to the typical matched-$\sigma$ ETG. However, this is not a fully satisfying explanation for several reasons. First, there is no known reason for the SNELLS lenses to differ systematically from matched-$\sigma$ ETGs. They clearly have atypical total $M/L$ values, which are compatible with a lighter IMF, but this origin of this difference is not understood, and the lenses are not found to be atypical in any other parameter. Second, there are some systematic uncertainties in the methods of IMF estimation that are not fully explored. For the stellar dynamical method (e.g., \ATD), it is important to understand how the inferred IMF varies if the stellar mass density profile does not follow the luminosity density, as is usually assumed, since significant differences between the two will arise when the IMF varies radially. For the SPS approach, although the global IMF trends appear to be robust, the normalization of $M_*/L$ depends on the spectral features used and the SPS model \citep[e.g.,][]{CvD12b,Spiniello15b}.

In particular, if we suppose that the treatment of sodium is in error and exclude the IMF-sensitive \ion{Na}{1} 8190~\AA~feature from our fits, we show in Appendix~B that the inferred $\alpha$ are significantly reduced. This provides an alternate way to reconcile the spectroscopic and lensing/dynamical IMF estimates within the SNELLS sample. However, it does so at the price of breaking the agreement between the spectroscopic $\alpha$ inferred from the SDSS stack and the lensing/dynamical $\alpha$ found by \ATD~and SLACS. While perhaps more acceptable than inconsistencies within SNELLS, given that the samples and apertures are not as precisely matched, we conclude that the exclusion of \ion{Na}{1} 8190~\AA~also does not provide a complete explanation, since it does not reconcile IMF estimates for both the SNELLS sample and the typical ETG with present data. Further comparisons of IMF estimates derived using different SPS codes, fitting methods, and spectral features would be a worthwhile future step toward establishing the robustness of these constraints. The SNELLS lenses are a good basis for such comparisons, and for this purpose, we provide the spectra used in this paper in Table~\ref{tab:data}.

\section{Summary}

We have compared lensing, dynamical, and SPS-based methods of estimating the IMF in the three low-redshift strong lenses located in the SNELLS survey. All were shown by \citet{Smith15} to have a relatively lightweight IMF based on their lensing masses, in contrast to the heavier IMFs claimed to be typical of similarly high-$\sigma$ ETGs based on earlier studies. We have investigated whether this discrepancy arises from scatter in the IMF of high-$\sigma$ ETGs or from a fundamental inconsistency among the methods. To do so, we analyzed new spectroscopic observations using new SPS models. Our principal findings are as follows.

\begin{enumerate}
%\item We confirm a lightweight IMF for the SNELLS lenses: $\langle \alpha_{\rm L+EAGLE} \rangle = 1.07 \pm 0.06$ as derived from lensing with EAGLE simulation-based estimates of the dark matter content, or alternatively $\langle \alpha_{\rm L+D} \rangle = 1.21 \pm 0.13$ from a joint lensing+dynamics analysis of SNL-0 and SNL-2.

\item Comparing lensing/dynamical estimates of the IMF mass factor $\alpha$ to SPS-based estimates derived from full spectral fitting, we find that all methods are consistent, regardless of the modeling assumptions, for one lens (SNL-2). For the other two lenses, the comparison depends on the dark matter content and the IMF parameterization used in the modeling. 

\item For IMFs with one or two power-law segments extending over $0.08$-$1\msol$, the spectroscopic $\alpha$ exceeds the lensing mass for SNL-0 and SNL-1. The significance depends on the dark matter fraction, but is 1.9-$2.5\sigma$ even if no dark matter is assumed. 
%Furthermore, for the same IMF parameterizations, the mean $\alpha$ of the SNELLS lenses matches that inferred from an SDSS composite spectrum of matched-$\sigma$ ETGs. This is inconsistent with the large difference in $\alpha$ between the SNELLS lenses and the mean matched-$\sigma$ ETG when lensing/dynamical measures are compared.

\item When adopting IMF parameterizations with more flexibility at low stellar masses, the tension among methods within the SNELLS sample decreases to 1.5$\sigma$ (SNL-0) and 1.9$\sigma$ (SNL-1) for a fiducial estimate of the dark matter content (based on the EAGLE simulations) and to $<1\sigma$ in the limit without dark matter. Thus, in the context of the CvD models, consistency at the $<2\sigma$ level among all methods of IMF estimation requires that the SNELLS galaxies, on average, have an elevated low-mass cutoff in the IMF or a turnover at low stellar masses $m \lesssim 0.3\msol$.

\item The IMFs of the SNELLS lenses are systematically lighter than the mean matched-$\sigma$ ETG as inferred by the \ATD~and SLACS surveys. This difference is evident in the lensing masses regardless of the dark matter fraction. If we instead adopt the total dynamical mass \emph{and} assume that no dark matter is present, the $M_*/L$ of the SNELLS lenses would be typical, but we argue that with current data the dynamical constraints are less precise and robust. This suggests that there may be substantial scatter in the IMF among high-$\sigma$ galaxies. 

\item The mean spectrum of the SNELLS lenses is very similar to an SDSS composite spectrum of matched-$\sigma$ ETGs, but subtle differences are detected at moderate statistical significance in the SPS analysis. If taken at face value, the spectral modeling suggests a deficit of low-mass stars in the SNELLS sample (either via an elevated low-mass cutoff, or a turnover at low masses in our non-parameteric models) but not for the SDSS composite. In addition, a truncated IMF is \emph{not} needed to reconcile spectroscopic and lensing/dynamical IMF estimates for the typical matched-$\sigma$ ETG, unlike the SNELLS lenses. Therefore, variation in form of the IMF at low stellar masses provides a plausible origin of the different lensing/dynamical IMF estimates of the two samples.

\item The SNELLS galaxies do not differ systematically in size, luminosity, mass density, metallicity, or abundance pattern from the mean matched-$\sigma$ ETG, so any scatter in the IMF that may be present does not appear to correlate strongly with global galaxy properties.

\item The absolute $M_*/L$ inferred from SPS modeling depends on the constraints used, particularly the inclusion of the \ion{Na}{1} 8190~\AA~feature. Omitting this constraint could reconcile IMF estimates within the SNELLS sample, but at the price of breaking the consistency found for the typical ETG.

\item We provide fully reduced spectra of the SNELLS lenses in order to facilitate future comparisons among SPS models. 

\end{enumerate}

\acknowledgements
We thank the referee for a thorough and thoughtful report that improved this work. We thank Lindsay Oldham for discussing unpublished observations. This paper includes data gathered with the 6.5 meter Magellan Telescopes located at Las Campanas Observatory, Chile. The FORS2 and X-shooter datasets used in this paper are available through the ESO data archives (program 095.B-0736). R.J.S.~acknowledges support from the STFC through grant ST/L00075X/1. C.C.~acknowledges support from NASA grant NNX15AK14G, NSF grant AST-1313280, and the Packard Foundation. A.V.~acknowledges the support of the NSF Graduate Research Fellowship.

\bibliographystyle{apj}
\bibliography{snells}

\appendix
\section{Comparison of 6dF and Literature Velocity Dispersions}

\begin{figure}
\centering
\includegraphics[width=0.5\linewidth]{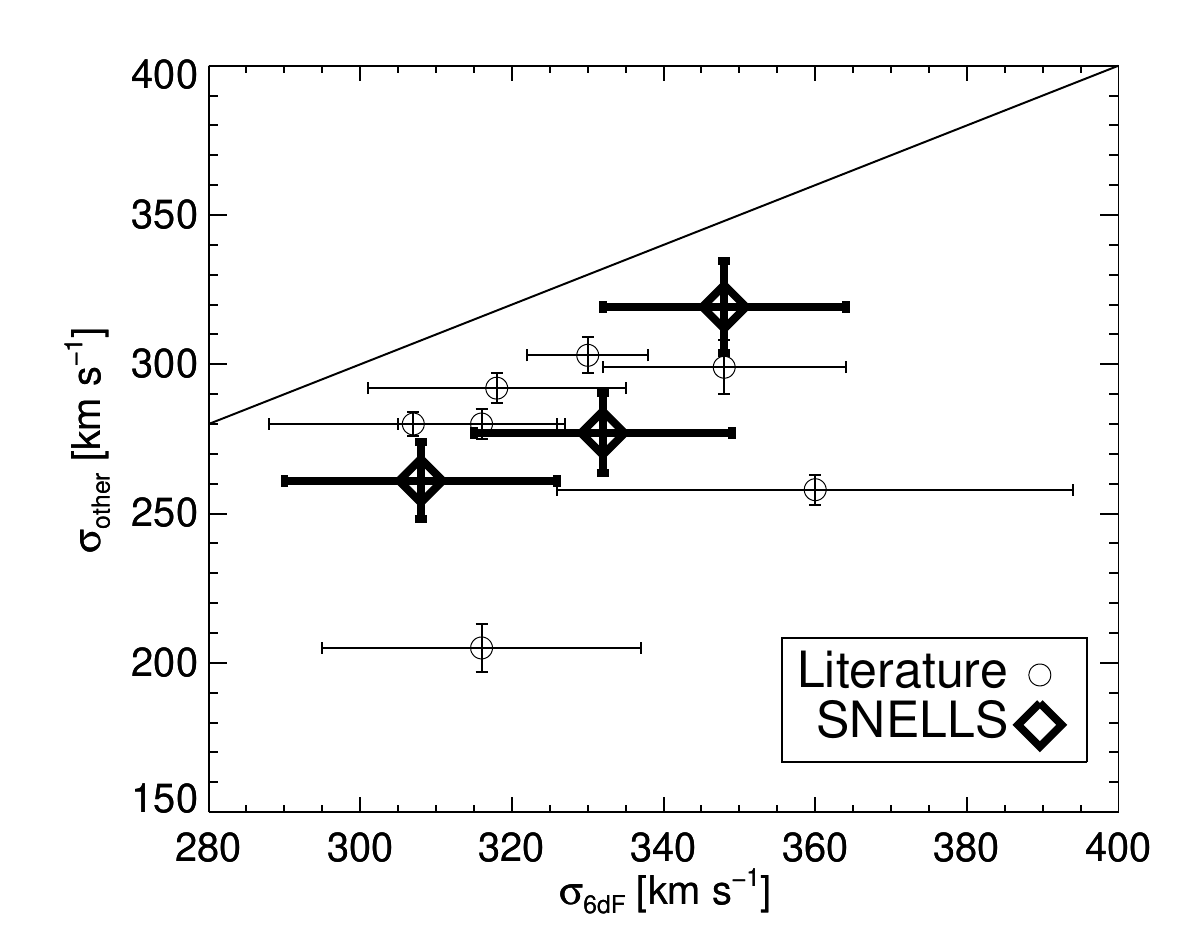}
\caption{Velocity dispersions of galaxies selected from the 6dF catalog with $\sigma_{\rm 6dF} > 300$~km~s${}^{-1}$ are compared to the measurements from this paper (diamonds) and from the literature sources (circles) described in the text. All data are corrected to a $6\farcs7$ diameter aperture.\label{fig:comparetolit}}
\end{figure}

As described in Section~\ref{sec:aperturedispersion}, the velocity dispersions measured in this paper are systematically lower than those in the 6dF catalog \citep{Campbell14} by $\sim 40$~km~s${}^{-1}$. Here we investigate this discrepancy further. Our measurements are based on new observations with $\simeq10-20\times$ higher signal-to-noise ratio than the 6dF spectra, and we obtain consistent $\sigma$ estimates from both the IMACS and X-shooter data using multiple fitting codes and templates. We are therefore confident in the new measurements and attribute the differences to a bias affecting the 6dF dispersions. To demonstrate this, we follow the selection criteria used by S15 to define the SNELLS parent sample. Specifically, we choose systems with $\sigma_{\rm 6dF} > 300$~km~s${}^{-1}$ and a signal-to-noise ratio of at least 15~\AA${}^{-1}$ in their 6dF spectrum. (We make no cut on environment for this test.) These criteria identify only 49 systems out of the 11315 galaxies with measured velocity dispersions in the \citet{Campbell14} catalog, i.e., the most extreme 0.4\% of the $\sigma_{\rm 6dF}$ distribution. Of these, we have identified seven galaxies with independent $\sigma$ measurements from the SDSS DR13 \citep{Albareti16}, \citet{Smith00}, or \citet{Jorgensen95}. These are compared to the 6dF dispersions in Figure~\ref{fig:comparetolit}, after standardizing to the $6\farcs7$ diameter 6dF aperture using the correction described in Section~\ref{sec:aperturedispersion}. In all cases, $\sigma_{\rm 6dF}$ exceeds the literature value. Furthermore, the mean excess $\langle \sigma_{\rm 6dF} - \sigma_{\rm lit} \rangle = 54\pm14$~km~s${}^{-1}$ is consistent with the mean difference between the 6dF measurements of the SNELLS lenses and those presented in this paper, $\langle \sigma_{\rm 6dF} - \sigma_{\rm SNELLS} \rangle = 44$~km~s${}^{-1}$. We conclude that the differences between our $\sigma$ measurements and the 6dF values used by S15 are typical, and that they reflect a bias that arises when selecting galaxies in the tail of the $\sigma$ distribution of the 6dF catalog. Although it is beyond the scope of this paper to examine this bias in detail, we note that it does not necessarily imply any error in the 6dF catalogs, since part or all of the observed effect must be the Eddington bias (see Section~\ref{sec:aperturedispersion}), which is purely statistical.

\section{Influence of \ion{Na}{1} on the Inferred IMF}

\begin{figure}
\centering
\includegraphics[width=0.6\linewidth]{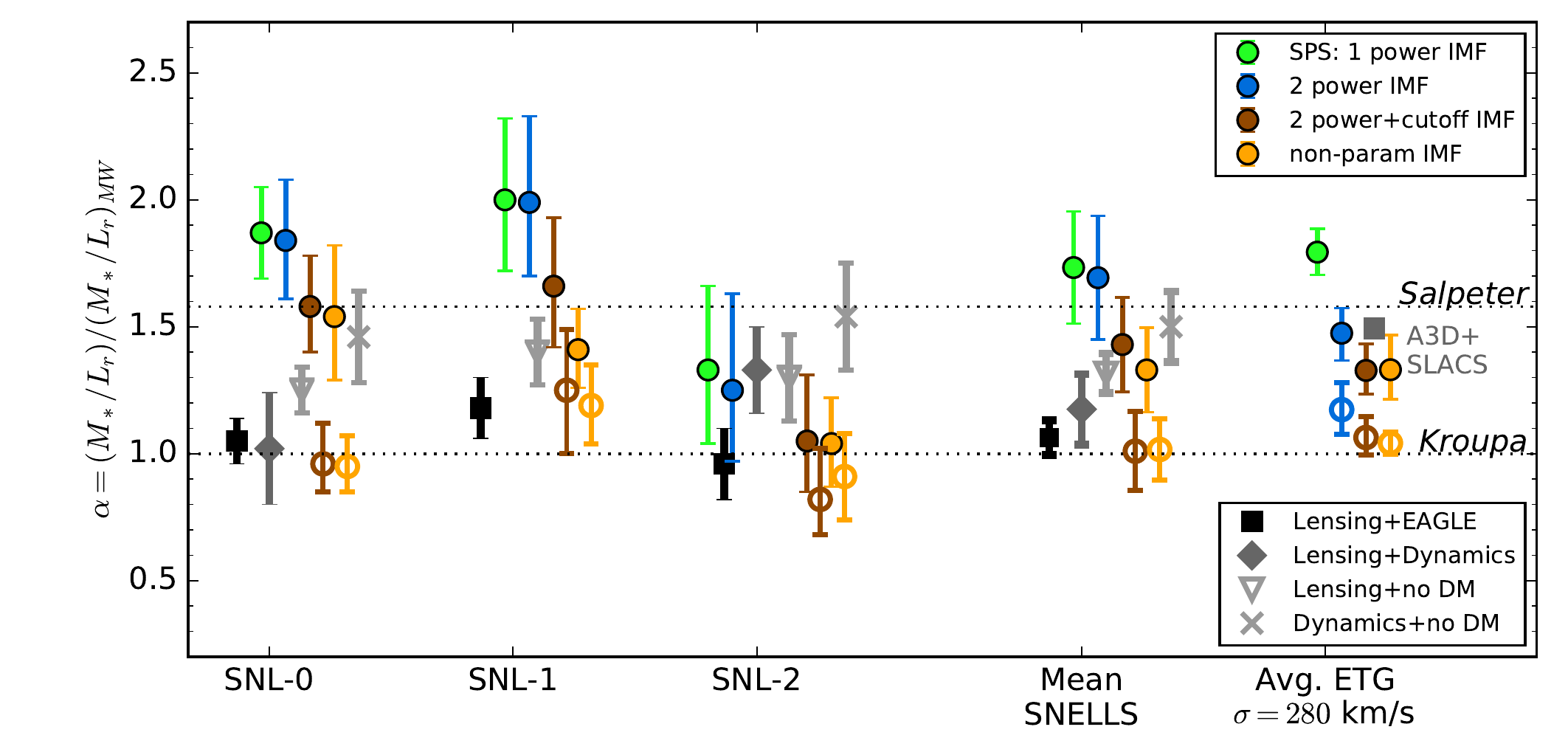}
\caption{Reproduction of Figure~\ref{fig:snells_sps}, with open circles added to represent the $\alpha$ inferred when the \ion{Na}{1} feature at 8190~\AA~is excluded from the fit.\label{fig:noNaI}}
\end{figure}

CvD12 showed that the normalization of the $\alpha$ values derived from their models is very sensitive to the inclusion of the \ion{Na}{1} 8190~\AA~feature in the fit. \citet{Smith15b} showed that these models, when constrained by the remainder of the optical-to-$J$-band spectrum, fail to match the strength of the \ion{Na}{1} $1.14\mu$m line. Similarly, \citet{Meneses-Goytia15} found that their models had difficulty reproducing the observed strength of the \ion{Na}{1} 2.21$\mu$m feature in massive ETGs. Recently, \citet{LaBarbera17} have claimed to fit the strength of all optical-NIR \ion{Na}{1} features consistently for the first time. Nonetheless, uncertainty about the accuracy of some models in reproducing the \ion{Na}{1} features motivates an examination of the influence of sodium in our IMF measurements.

Figure~\ref{fig:noNaI} reproduces Figure~\ref{fig:snells_sps} with additional open symbols that show the results obtained when the \ion{Na}{1} 8190~\AA~feature is excluded from the fit. Excluding \ion{Na}{1} 8190~\AA~reduces the inferred $\alpha$ by $\simeq 30\%$ on average, consistent with findings by CvD12. Such a reduction is sufficient to remove all tension between lensing/dynamical and spectroscopic IMF constraints in the SNELLS sample. However, as seen in this figure and discussed in Section~\ref{sec:disc}, masking \ion{Na}{1} introduces substantial tension among these IMF constraints when the typical matched-$\sigma$ ETG is considered. Regardless of the IMF parameterization, the SPS-derived $\alpha$ values fall significantly below the lensing/dynamics constraints when \ion{Na}{1} 8190~\AA~is masked. 
The exclusion of this feature due to potential systematic errors mainly rescales all SPS-derived $M_*/L$ downward uniformly. Therefore, it cannot reproduce the separation seen between the SNELLS galaxies and the SDSS stack in their lensing/dynamics-based $\alpha$, which is the main subject of this paper. As an aside, we note that although masking \ion{Na}{1} 8190~\AA~pushes the SDSS stack to Kroupa-like $M/L$ values, this does not necessarily eliminate the spectroscopic evidence for systematic IMF variations: this question depends on how the inference changes for lower-$\sigma$ galaxies, which is beyond the scope of this paper. 

\end{document}